\documentclass[twoside,twocolumn,9pt]{article}
\usepackage{extsizes}
\usepackage[super,sort&compress,comma]{natbib} 
\usepackage[version=3]{mhchem}
\usepackage[left=1.5cm, right=1.5cm, top=1.785cm, bottom=2.0cm]{geometry}
\usepackage{balance}
\usepackage{mathptmx}
\usepackage{sectsty}
\usepackage{graphicx} 
\usepackage{lastpage}
\usepackage[format=plain,justification=justified,singlelinecheck=false,font={stretch=1.125,small,sf},labelfont=bf,labelsep=space]{caption}
\usepackage{float}
\usepackage{fancyhdr}
\usepackage{fnpos}
\usepackage[english]{babel}
\addto{\captionsenglish}{%
  
}
\usepackage{array}
\usepackage{droidsans}
\usepackage{charter}
\usepackage[T1]{fontenc}
\usepackage[usenames,dvipsnames]{xcolor}
\usepackage{setspace}
\usepackage[compact]{titlesec}
\usepackage{hyperref}

\usepackage{epstopdf}
\usepackage{enumitem}

\definecolor{cream}{RGB}{222,217,201}

\begin{document}

\pagestyle{fancy}
\thispagestyle{plain}
\fancypagestyle{plain}{
\renewcommand{\headrulewidth}{0pt}
}

\makeFNbottom
\makeatletter
\renewcommand\LARGE{\@setfontsize\LARGE{15pt}{17}}
\renewcommand\Large{\@setfontsize\Large{12pt}{14}}
\renewcommand\large{\@setfontsize\large{10pt}{12}}
\renewcommand\footnotesize{\@setfontsize\footnotesize{7pt}{10}}
\makeatother

\renewcommand{\thefootnote}{\fnsymbol{footnote}}
\renewcommand\footnoterule{\vspace*{1pt}%
\color{cream}\hrule width 3.5in height 0.4pt \color{black}\vspace*{5pt}} 
\setcounter{secnumdepth}{5}

\makeatletter 
\renewcommand\@biblabel[1]{#1}            
\renewcommand\@makefntext[1]%
{\noindent\makebox[0pt][r]{\@thefnmark\,}#1}
\makeatother 
\renewcommand{\figurename}{\small{Fig.}~}
\sectionfont{\sffamily\Large}
\subsectionfont{\normalsize}
\subsubsectionfont{\bf}
\setstretch{1.125} 
\setlength{\skip\footins}{0.8cm}
\setlength{\footnotesep}{0.25cm}
\setlength{\jot}{10pt}
\titlespacing*{\section}{0pt}{4pt}{4pt}
\titlespacing*{\subsection}{0pt}{15pt}{1pt}

\fancyfoot{}
\fancyfoot[LO,RE]{\vspace{-7.1pt}\includegraphics[height=9pt]{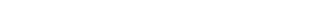}}
\fancyfoot[CO]{\vspace{-7.1pt}\hspace{13.2cm}\includegraphics{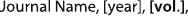}}
\fancyfoot[CE]{\vspace{-7.2pt}\hspace{-14.2cm}\includegraphics{head_foot/RF}}
\fancyfoot[RO]{\footnotesize{\sffamily{1--\pageref{LastPage} ~\textbar  \hspace{2pt}\thepage}}}
\fancyfoot[LE]{\footnotesize{\sffamily{\thepage~\textbar\hspace{3.45cm} 1--\pageref{LastPage}}}}
\fancyhead{}
\renewcommand{\headrulewidth}{0pt} 
\renewcommand{\footrulewidth}{0pt}
\setlength{\arrayrulewidth}{1pt}
\setlength{\columnsep}{6.5mm}
\setlength\bibsep{1pt}

\makeatletter 
\newlength{\figrulesep} 
\setlength{\figrulesep}{0.5\textfloatsep} 

\newcommand{\topfigrule}{\vspace*{-1pt}%
\noindent{\color{cream}\rule[-\figrulesep]{\columnwidth}{1.5pt}} }

\newcommand{\botfigrule}{\vspace*{-2pt}%
\noindent{\color{cream}\rule[\figrulesep]{\columnwidth}{1.5pt}} }

\newcommand{\dblfigrule}{\vspace*{-1pt}%
\noindent{\color{cream}\rule[-\figrulesep]{\textwidth}{1.5pt}} }

\makeatother

\twocolumn[
  \begin{@twocolumnfalse}
{\includegraphics[height=30pt]{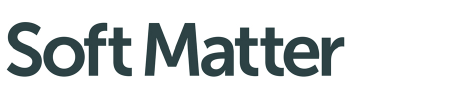}\hfill\raisebox{0pt}[0pt][0pt]{\includegraphics[height=55pt]{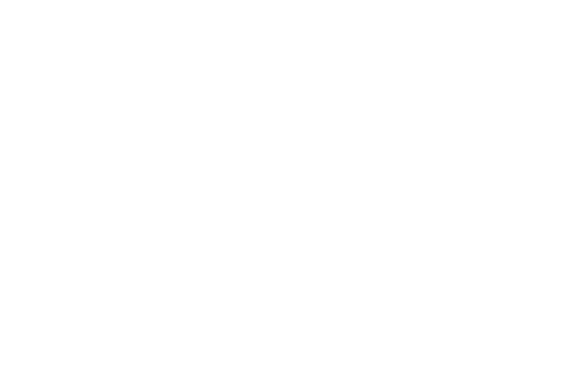}}\\[1ex]
\includegraphics[width=18.5cm]{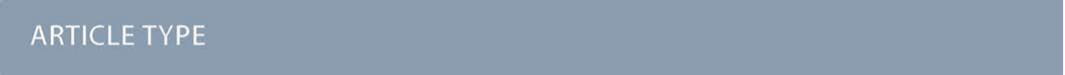}}\par
\vspace{1em}
\sffamily
\begin{tabular}{m{4.5cm} p{13.5cm} }

\includegraphics{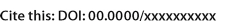} & \noindent\LARGE{\textbf{Interplay between hysteresis and nonlocality during onset and arrest of flow in granular materials}} \\
\vspace{0.3cm} & \vspace{0.3cm} \\

 & \noindent\large{Saviz Mowlavi\textit{$^{a}$} and Ken Kamrin\textit{$^{a}$}} \\

\includegraphics{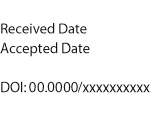} & \noindent\normalsize{The jamming transition in granular materials is well-known for exhibiting hysteresis, wherein the level of shear stress required to trigger flow is larger than that below which flow stops. Although such behavior is typically modeled as a simple non-monotonic flow rule, the rheology of granular materials is also nonlocal due to cooperativity at the grain scale, leading for instance to increased strengthening of the flow threshold as system size is reduced. We investigate how these two effects -- hysteresis and nonlocality -- couple with each other by incorporating non-monotonicity of the flow rule into the nonlocal granular fluidity (NGF) model, a nonlocal constitutive model for granular flows. By artificially tuning the strength of nonlocal diffusion, we demonstrate that both ingredients are key to explaining certain features of the hysteretic transition between flow and arrest. 
{Finally, we assess the ability of the NGF model to quantitatively predict material behavior both around the transition and in the flowing regime, through stress-driven discrete element method (DEM) simulations of flow onset and arrest in various geometries.} 
Along the way, we develop a new methodology to compare deterministic model predictions with the stochastic behavior exhibited by the DEM simulations around the jamming transition.} \\
\end{tabular}
\end{@twocolumnfalse} \vspace{0.6cm}]

\renewcommand*\rmdefault{bch}\normalfont\upshape
\rmfamily
\section*{}
\vspace{-1cm}


\footnotetext{\textit{$^{a}$~Department of Mechanical Engineering, Massachusetts Institute of Technology, Cambridge, MA 02139, USA; E-mail: kkamrin@mit.edu}}



\section{Introduction}

\begin{figure*}[htb]
 \centering
 \includegraphics[width=\textwidth]{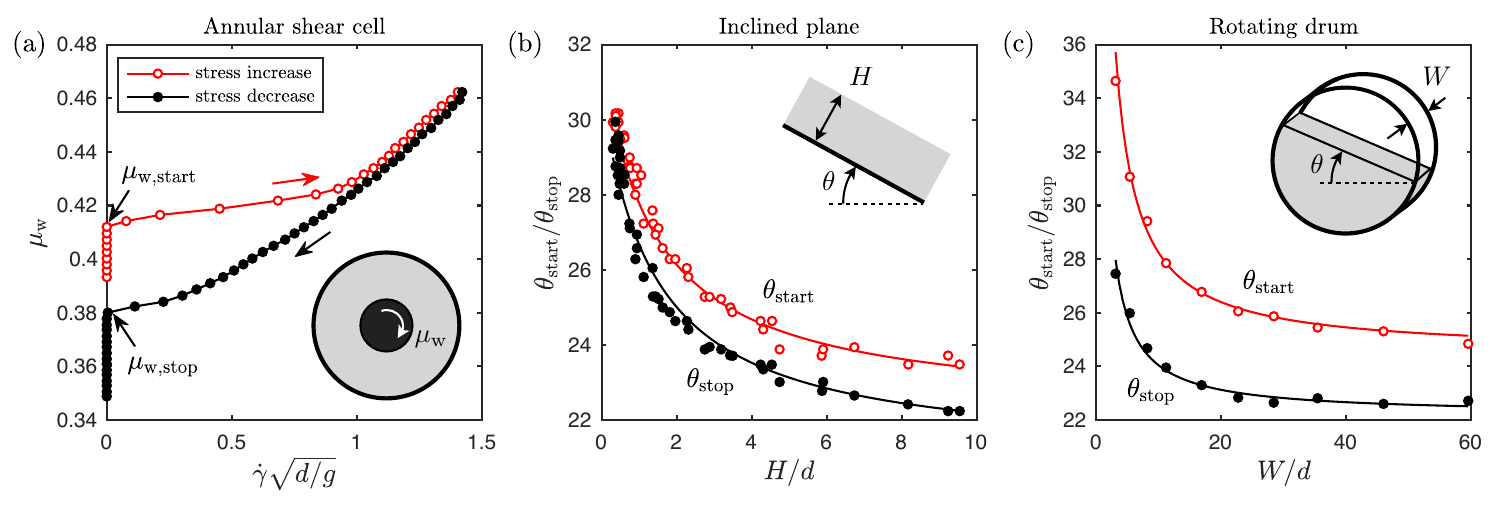}
 \caption{Previous experimental investigations of the flow threshold in various geometries. (a) Annular shear cell \cite{dacruz2002}: ratio of shear stress to pressure at the inner wall, $\mu_\mathrm{w}$, versus dimensionless mean strain rate, $\dot{\gamma} \sqrt{d/g}$, for increasing and decreasing torque applied to the inner cylinder. (b) Inclined plane \cite{pouliquen2002}: angles of inclination at flow onset and arrest, $\theta_\mathrm{start}$ and $\theta_\mathrm{stop}$, versus dimensionless layer thickness, $H/d$. (c) Rotating drum \cite{courrech2003b}: angles of inclination at flow onset and arrest, $\theta_\mathrm{start}$ and $\theta_\mathrm{stop}$, versus dimensionless drum width, $W/d$.}
 \label{fig:Experiments}
\end{figure*}

Granular materials are well-known for displaying both solid-like and fluid-like behavior depending on their internal stress state \citep{forterre2008,andreotti2013,srivastava2021}. Flow can be induced or arrested through external loading variations, which has direct implications for a wide range of catastrophic geophysical phenomena such as landslides, avalanches and earthquakes \cite{daerr1999,lucas2014,scholz1998}. The transition between solid-like and liquid-like behavior in frictional granular media is characterized by several unique macroscopic features, which have been uncovered through simple experiments in model systems \cite{midi2004}. Figure \ref{fig:Experiments} showcases typical results from such experiments, where flow is triggered then arrested by ramping up and down the applied stress in (a) an annular shear cell \cite{dacruz2002}, (b) a layer of grains on an inclined plane \cite{pouliquen2002}, and (c) a partially-filled rotating drum \cite{courrech2003b}.
The features revealed in these experiments are universal to most geometries and can be outlined as follows: 
\begin{enumerate}[label=(F\arabic*),leftmargin=25pt,itemsep=-2pt]
\item the level of stress required to trigger flow is larger than that below which flow stops, leading to a hysteresis of the flow velocity as the applied stress is ramped up and down;
\item the onset of flow is accompanied by a finite jump in the velocity of the system;
\item the critical stresses for onset and arrest of flow depend on the size\footnote{By system size, we refer to the relevant length scale controlling the width of shear regions in the flow field. Depending on the geometry, this length scale can either be geometric or stress-induced.} of the system, with smaller system sizes displaying increased strengthening.
\end{enumerate}
Each feature is directly relevant to geophysical events such as landslides and avalanches, since (F1) controls the mobilized mass that flows down, (F2) explains why they are so spontaneous and catastrophic, and (F3) determines the circumstances under which they might occur.
The objective of the present work is to formulate a continuum model that is able to describe quantitatively the onset and arrest of flow in frictional granular materials in various two-dimensional geometries, and analyze how its constituent ingredients play a role in reproducing each of these three features, with particular focus on geometries displaying inhomogeneous flow fields.

It is now well accepted that dense and homogeneous flows of grains follow the $\mu(I)$ constitutive relationship, which states that the stress ratio $\mu$ and the inertial number $I$ are related through a one-to-one function $\mu = \mu_\text{loc}(I)$ \cite{midi2004,dacruz2005}. In two dimensions, $\mu = \tau/P$ is the ratio of shear stress $\tau$ to pressure $P$, and $I = \dot{\gamma} \sqrt{m/P}$ is the strain rate $\dot{\gamma}$ nondimensionalized with a particle-wise rearrangement time scale formed by the mean grain mass $m$ and confining pressure $P$. While $\mu_\text{loc}(I)$ has long been believed to be a monotonic function of $I$, several recent experiments \cite{dijksman2011,kuwano2013,perrin2019,russell2019} and simulations \cite{yang2016,degiuli2017} have revealed the existence at very low $I$ of a strain-rate weakening regime, wherein the stress ratio $\mu$ decreases with increasing $I$. One possible non-monotonic functional form is
\begin{equation}
\mu_{\text{loc}}(I) = \mu_s + \frac{\mu_2-\mu_s}{(\mu_2-\mu_s)/(bI+\chi(I;\kappa))+1},
\label{eq:LocalRheology}
\end{equation}
where $\mu_s$, $\mu_2$, and $b$ are dimensionless rheological constants, $\kappa = k_n/P$ a dimensionless stiffness with $k_n$ the grain stiffness, and $\chi$ is a decreasing function of $I$ that accounts for the strain-rate weakening regime. The microscopic origin of the strain-rate weakening regime has recently come under debate, with some studies arguing that it is caused by inertia of the grains \cite{quartier2000,courrech2003,degiuli2017}, while others observing velocity-weakening behavior in over-damped, inertia-less particulate media \cite{perrin2019}. The present work is not concerned with this particular issue, and we simply leave open the possibility for the amount of strain-rate weakening to depend on the grain stiffness through the dimensionless parameter $\kappa$ entering $\chi$. In any case, the non-monotonicity of \eqref{eq:LocalRheology} necessarily implies that features (F1) and (F2) above are realized in homogeneous flows: the level of stress required to trigger flow is higher than that at which flow stops, and flow onset is characterized by a velocity jump\cite{jaeger1990,mills2008}.

However, the $\mu(I)$ constitutive relationship breaks down in inhomogeneous flows, in the sense that $\mu$ is no longer a one-to-one function of $I$ \cite{koval2009,tang2018}. Due to the finite size of the grains, velocity fluctuations generated at an arbitrary location will spread over some grain-size-dependent correlation length and change the rheology of the neighboring material \cite{melosh1979,goyon2008,reddy2011,gaume2011}, resulting in wider shear regions than are predicted by the $\mu(I)$ rheology, especially in the quasi-static limit \cite{midi2004}. Such spatial cooperativity at the scale of individual grains also explains why thinner layers on an inclined plane start flowing at higher inclination angles than thicker layers, despite the stress ratio $\mu$ being independent of the layer height \cite{midi2004}. Nonlocal rheological models, which incorporate an intrinsic length scale, have been shown to capture several of these phenomena \cite{kamrin2019}. Here, we focus on the nonlocal granular fluidity (NGF) model \cite{kamrin2012,henann2013}, which relates the stress ratio and strain rate through a granular fluidity field $g = \dot{\gamma}/\mu$ that is governed by a reaction-diffusion partial differential equation (PDE). Our choice of the NGF model stems from its ability to reproduce the system-size dependence of the flow threshold in various geometries \cite{kamrin2015,liu2018}, which partially explains feature (F3) above. But the current formulation of the NGF model reduces to the monotonic form of the $\mu_\text{loc}(I)$ relationship in homogeneous flow conditions, meaning that the model does not have a built-in mechanism to account for the remaining features (F1) and (F2). 

\begin{figure*}[htb]
 \centering
 \includegraphics[width=\textwidth]{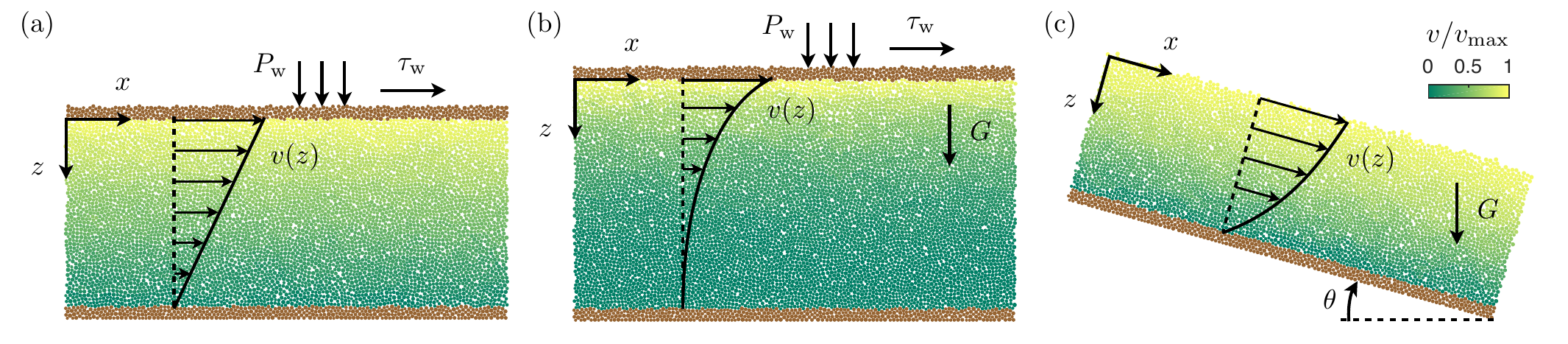}
 \caption{Geometries considered in this study: (a) plane simple shear, (b) plane shear under gravity, and (c) inclined plane flows. Particles that are free to flow are colored according to their relative velocity magnitude, and fixed wall particles are shown in brown.}
 \label{fig:Geometries}
\end{figure*}

In this paper, we modify the NGF model so that it instead reduces to the non-monotonic form of the $\mu_\text{loc}(I)$ relationship, equation \eqref{eq:LocalRheology}, in homogeneous flows. By computing time-dependent model predictions in a stress-driven planar shear configuration under gravity, we evaluate the specific ways in which nonlocality and non-monotonicity contribute to each of the three features (F1--F3) of the flow-arrest transition in inhomogeneous flows. We show that inclusion of, and interplay between both ingredients is necessary to reproduce all three features, in ways sometimes surprising: the planar shear with gravity configuration displays a finite velocity jump during onset of flow only when both non-monotonicity and nonlocality are present. In a second part, we assess the capability of the modified NGF model to predict quantitatively the behavior of dense granular materials both around the flow-arrest transition as well as in the flowing regime. To this effect, we calibrate the model using discrete element method (DEM) simulations in the simple shear geometry shown in Figure \ref{fig:Geometries}(a), and we compare predictions of the calibrated model against stress-driven DEM simulations in the other geometries displayed in Figures \ref{fig:Geometries}(b) and \ref{fig:Geometries}(c), namely plane shear under gravity and inclined plane.
These two configurations are both subject to nonlocal effects as a result of the spatial inhomogeneity of their flow fields, but they critically differ in an important way -- in plane shear with gravity, flow inhomogeneity is mostly a consequence of the spatial dependence of the stress ratio $\mu$, while it is mainly caused by the rough base in inclined plane flow \cite{liu2018}. We observe that the accuracy of the NGF model depends on which of these two mechanisms is at play, with predictions being accurate in the case of plane shear with gravity but less so for inclined plane{, which we ultimately attribute to the role played by the boundary conditions}.

The remainder of this paper is organized as follows. In Section \ref{sec:NonlocalGranularRheology}, we present a modified NGF model that incorporates a non-monotonic local rheology, and we evaluate the combined effects of nonlocality and non-monotonicity on the features of the flow-arrest transition. We then compare in Section 3 predictions from the NGF model with DEM simulation results in stress-driven planar shear with gravity and inclined plane configurations. We close the paper with concluding remarks in Section 4.

\section{Nonlocal granular rheology}
\label{sec:NonlocalGranularRheology}

In this section, we discuss our nonlocal continuum modelling approach based on the nonlocal granular fluidity (NGF) model. We begin by presenting the NGF model in its current form, which does not capture the hysteresis of the flow-arrest transition. We then describe the incorporation of bistable behavior into the NGF model. Finally, we evaluate the combined effects of bistability and nonlocality on the qualitative behavior of the flow-arrest transition.

\subsection{Nonlocal model without hysteresis}

Extending earlier fluidity-based nonlocal models for concentrated emulsions \cite{picard2005,goyon2008,bocquet2009} to granular materials, the NGF model introduces a positive granular fluidity field $g$ that relates the strain rate $\dot{\gamma}$ with the stress ratio $\mu$ through the following two constitutive equations:
\begin{subequations}
\begin{align}
\dot{\gamma} &= g \mu, \label{eq:NGF1} \\
t_0 \dot{g} &= A^2 d^2 \nabla^2 g - \frac{(\mu_2-\mu_\mathrm{s})(\mu_\mathrm{s}-\mu)}{\mu_2-\mu} g - b \sqrt{\frac{m}{P}} \mu g^2, \label{eq:NGF2}
\end{align} \label{eq:NGF}%
\end{subequations}%
where $t_0$ is a constant timescale associated with the dynamics of $g$, and the nonlocal amplitude $A > 0$ is a dimensionless scalar parameter quantifying the strength of spatial cooperativity in the flow. As a side note, we mention that recent studies \cite{zhang2017,bhateja2018,kim2020} have endowed the granular fluidity field with a clear physical meaning -- $g$ is a purely kinematic quantity related to the velocity fluctuations $\delta v$, grain size $d$ and solid fraction $\phi$ through $g = (\delta v/d) F(\phi)$, where $F$ depends solely on $\phi$.

The flow rule \eqref{eq:NGF1} states that the strain rate $\dot{\gamma}$ is directly proportional to the fluidity $g$. Therefore, there can only be flow provided $g$ is nonzero, and the nonlocal granular rheology is driven by the dynamics of equation \eqref{eq:NGF2} for the fluidity. The latter takes the form of a reaction-diffusion equation, and its behavior can be understood as follows.

\begin{figure*}
 \centering
 \includegraphics[width=\textwidth]{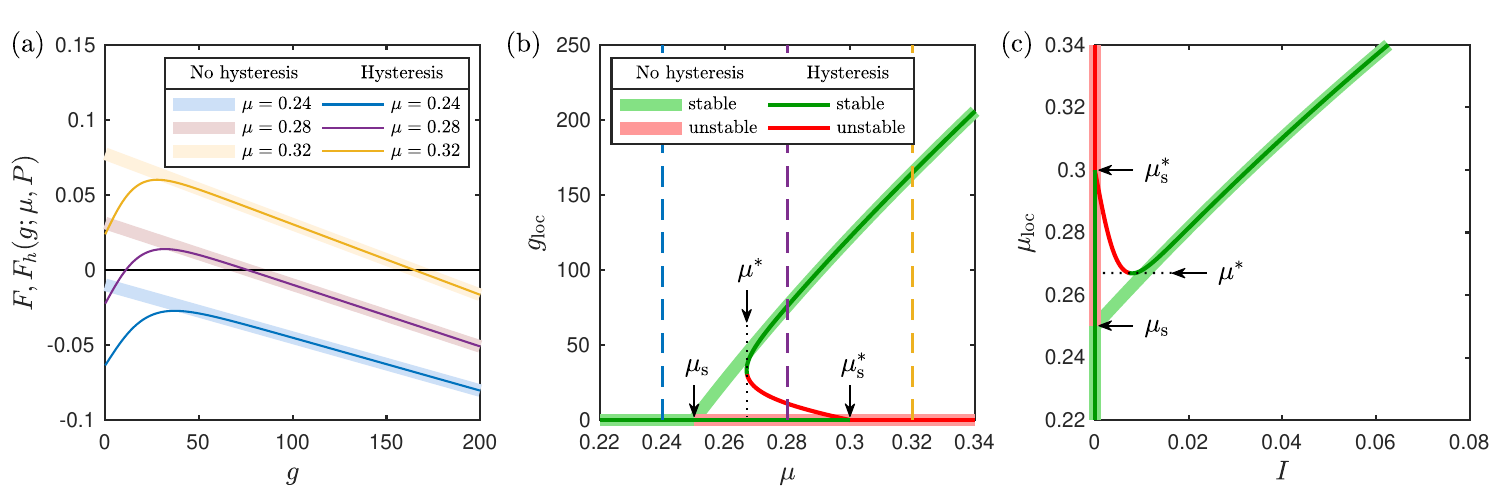}
 \caption{Steady-state solutions of the local limit of the NGF model for homogeneous flows, without hysteresis (thick lines) and with hysteresis (thin lines). (a) Behavior of $F(g;\mu,P)$ and $F_h(g;\mu,P)$ in \eqref{eq:NGF2_loc} and \eqref{eq:HysNGF2_loc}, showing the existence of zero, one or two roots for different stress ratios $\mu$. (b) Unstable (red) and stable (green) steady-state local solutions $g_\mathrm{loc}$ as a function of $\mu$. (c) Resultant local rheology $\mu_\mathrm{loc}(I)$, with both the unstable (red) and stable (green) branches shown. The model with hysteresis displays a strain-rate weakening regime absent in the model without hysteresis.}
 \label{fig:LocalRheology}
\end{figure*}

In the absence of boundary effects or nonuniformities in the stress ratio $\mu$, the $g$ field becomes spatially uniform and \eqref{eq:NGF2} reduces to the simple dynamical system
\begin{equation}
t_0 \dot{g} = - \frac{(\mu_2-\mu_\mathrm{s})(\mu_\mathrm{s}-\mu)}{\mu_2-\mu} g - b \sqrt{\frac{m}{P}} \mu g^2 \equiv F(g;\mu,P) g,
\label{eq:NGF2_loc}
\end{equation}
where $F(g;\mu,P)$ is a simple linear function of $g$. The steady behavior of the system is then governed by the steady-state solutions $g_\mathrm{loc}$ of \eqref{eq:NGF2_loc}, which we illustrate by the thick lines in Figure \ref{fig:LocalRheology} using arbitrary values\footnote{We note that the qualitative behavior of the model is independent of its specific parameter values. The latter are therefore chosen to be reasonably close to the calibrated values obtained later in Section \ref{sec:CalibrationSimplePlaneShear}, while displaying the hysteretic behavior of the model with enough clarity in Figures \ref{fig:LocalRheology} and \ref{fig:Schema}.} for $\mu_\mathrm{s} = 0.25$, $\mu_2$, $b$, and $P$.
When $\mu \le \mu_\mathrm{s}$, the function $F(g;\mu,P) < 0$ for all $g \ge 0$ as shown in Figure \ref{fig:LocalRheology}(a), hence \eqref{eq:NGF2_loc} only admits the stable, arrested steady state $g_\mathrm{loc} = 0$, represented by the zero green branch in Figure \ref{fig:LocalRheology}(b). When $\mu > \mu_\mathrm{s}$, the linear function $F(g;\mu,P)$ acquires one positive root and $F(g=0;\mu,P) > 0$, which has two consequences. First, the arrested steady state becomes unstable, shown by the red branch in Figure \ref{fig:LocalRheology}(b). Second, a stable, flowing steady-state solution $g_\mathrm{loc}(\mu) > 0$ emerges, represented by the positive green branch in Figure \ref{fig:LocalRheology}(b). Using the flow rule \eqref{eq:NGF1} together with the definition of the inertial number $I$, the flowing and arrested stable solutions can be inverted and expressed as
\begin{equation}
\mu_\mathrm{loc}(I) = \mu_s + \frac{\mu_2-\mu_s}{(\mu_2-\mu_s)/bI+1},
\label{eq:LocalRheologyMonotonic}
\end{equation}
for $I > 0$, and $\mu_\mathrm{loc} \le \mu_\mathrm{s}$ otherwise. This relationship is pictured in green in Figure \ref{fig:LocalRheology}(c), together with the unstable arrested solution above $\mu_s$ in red. Therefore, the NGF model reduces to the local $\mu(I)$ rheology in steady and homogeneous flows such as plane shear without gravity. 

In the presence of boundary effects or nonuniformities in the stress ratio $\mu$, however, the diffusion term in \eqref{eq:NGF2} spreads the granular fluidity over a cooperativity length scale proportional to the grain size $d$, resulting in a nonlocal flow rule \eqref{eq:NGF1}. Regions where $\mu > \mu_\mathrm{s}$ act as stress-driven sources of granular fluidity, which is then diffused towards lower-stress regions or boundaries. Such nonlocal, cooperative effects have manifold consequences, and the NGF model explains many phenomena evading local rheological models. For instance, the model recovers the decaying motion of grains in regions where $\mu < \mu_\mathrm{s}$ \cite{henann2013,liu2017,tang2018} as well as the so-called secondary rheology, wherein flow anywhere in a granular media removes the yield stress elsewhere \cite{henann2014,li2020}. Conversely, the model is able to explain the strengthening of the flow threshold with decreasing system size \cite{kamrin2015,liu2018}, which is caused by boundaries or low-stress regions preventing flow in other regions where $\mu > \mu_\mathrm{s}$ unless $\mu$ is large enough. This last property relates to feature (F3) mentioned in the introduction. Yet, the current form of the NGF model is unable to reproduce features (F1) and (F2) due to the monotonicity of its limiting local rheology \eqref{eq:LocalRheologyMonotonic}.

\subsection{Nonlocal model with hysteresis}

We now discuss the inclusion of non-monotonicity of the local rheological response into the NGF model. Taking inspiration from previous hysteretic nonlocal models \cite{aranson2002,lee2017}, we add a new term to the right-hand side of the fluidity equation \eqref{eq:NGF2}. The constitutive equations \eqref{eq:NGF} become
\begin{subequations}
\begin{align}
\dot{\gamma} &= g \mu, \label{eq:HysNGF1} \\
t_0 \dot{g} &= A^2 d^2 \nabla^2 g - \frac{(\mu_2-\mu_\mathrm{s})(\mu_\mathrm{s}-\mu)}{\mu_2-\mu} g - b \sqrt{\frac{m}{P}} \mu g^2 \nonumber \\
&\quad - \chi(g; \mu, P) g, \label{eq:HysNGF2}
\end{align} \label{eq:HysNGF}%
\end{subequations}%
where the new term $\chi(g; \mu, P)$ takes the form
\begin{equation}
\chi(g; \mu, P) = a \left[1-\tanh \left(c \sqrt{\frac{m}{P}} \mu g \kappa^n \right) \right],
\end{equation}
with $a$, $c$, $n$ constant scalar parameters, and $\kappa = k_n/P$ the nondimensional particle stiffness. Here, we choose to express the new term $\chi$ with a $\tanh$ function so that it vanishes for large $g$, which restricts its contribution to the behavior of the system near the jamming transition.

As before, we begin by evaluating the dynamics of the $g$ field for the case of homogeneous flows, in which $g$ is spatially uniform and \eqref{eq:HysNGF2} reduces to the simple dynamical system
\begin{align}
t_0 \dot{g} &= - \frac{(\mu_2-\mu_\mathrm{s})(\mu_\mathrm{s}-\mu)}{\mu_2-\mu} g - b \sqrt{\frac{m}{P}} \mu g^2 - \chi(g; \mu, P) g \nonumber \\
&\equiv F_h(g;\mu,P) g,  \label{eq:HysNGF2_loc}
\end{align}
where $F_h(g;\mu,P)$ is a function of $g$. Contrary to the previous case without hysteresis, the presence of the $\chi$ term induces a decrease in $F_h(g;\mu,P)$ when $g$ approaches zero, as illustrated by the thin lines in Figure \ref{fig:LocalRheology}(a) using the same parameter values as before and new arbitrary values for the parameters of $\chi$. As a result, there exists a range of stress ratios $\mu^* < \mu < \mu_\mathrm{s}^*$ in which the function $F_h(g;\mu,P)$ inherits a second positive root, so that \eqref{eq:HysNGF2_loc} admits two flowing steady-state solutions $g_\mathrm{loc}(\mu) > 0$, one stable and one unstable, shown by the thin lines in Figure \ref{fig:LocalRheology}(b). These two flowing solution branches merge at $\mu = \mu^*$. The stable branch reverts for $\mu > \mu_\mathrm{s}^*$ to the same flowing solution as the NGF model without hysteresis,
while the unstable branch merges at $\mu = \mu_\mathrm{s}^*$ with the arrested steady-state solution $g_\mathrm{loc} = 0$, which remains stable until $\mu$ exceeds $\mu_\mathrm{s}^*$.
Collecting the pieces, the steady-state solutions of \eqref{eq:HysNGF2_loc}
 can be expressed in terms of the inertial number $I$ as
\begin{equation}
\mu_\mathrm{loc}(I) = \mu_s + \frac{\mu_2-\mu_s}{(\mu_2-\mu_s)/(bI+\chi(I; \kappa))+1},
\label{eq:LocalRheologyNonmonotonic}
\end{equation}
for $I > 0$, and $\mu_\mathrm{loc} \le \mu_\mathrm{s}^*$ otherwise. The function $\chi$ is now formulated in terms of $I$ and $\kappa$ as
\begin{equation}
\chi(I; \kappa) = a \left[1-\tanh \left(c I \kappa^n \right) \right],
\end{equation}
and the static yield stress ratio $\mu_\mathrm{s}^*$ is obtained as
\begin{equation}
\mu_\mathrm{s}^* = \mu_\mathrm{loc}(I \rightarrow 0) = \frac{\mu_\mathrm{s}(\mu_2-\mu_\mathrm{s}) + a \mu_2}{\mu_2 - \mu_\mathrm{s} + a}.
\label{eq:YieldStressNonmonotonic}
\end{equation}
Thus, the modified NGF model reduces to the non-monotonic $\mu(I)$ rheology \eqref{eq:LocalRheology} in steady homogeneous flows, and the corresponding stable and unstable branches are displayed by the thin green and red lines in Figure \ref{fig:LocalRheology}(c). In the quasi-static, low-$I$ regime, the presence of $\chi$ induces a weakening relationship between $\mu$ and $I$. For higher values of $I$, the vanishing of $\chi$ leads to a strain-rate strengthening regime in which the non-monotonic local rheology \eqref{eq:LocalRheologyNonmonotonic} converges to its monotonic counterpart \eqref{eq:LocalRheologyMonotonic}. The crossover between the two regimes occurs at $d\mu_\mathrm{loc}/dI = 0$, which corresponds to
\begin{equation}
I^* = \frac{1}{c \kappa^n} \mathrm{sech}^{-1} \sqrt{\frac{b}{ac\kappa^n}} \quad \text{and} \quad \mu^* = \mu_\mathrm{loc}(I^*).
\end{equation}

In agreement with force balance arguments \cite{jaeger1990}, the strain-rate weakening regime is unstable while the strain-rate strengthening regime is stable. Since the latter exists for $\mu$ above $\mu^*$ and the arrested solution is stable below $\mu_\mathrm{s}^*$, there exist two stable steady-state solutions -- one flowing and one arrested -- in the range $\mu^* < \mu < \mu_\mathrm{s}^*$. In the absence of flow gradients, this bistable behavior generates hysteresis when the stress ratio $\mu$ is ramped up and down: flow is triggered at $\mu_\mathrm{s}^*$ but stops at a lower $\mu^*$. In addition, the onset of flow is accompanied by a finite jump in the velocity of the system, as the inertial number jumps from the arrested solution to the stable flowing solution. Hence features (F1) and (F2) are accounted for, but it is unclear whether this would hold for inhomogeneous flow, in the presence of nonlocal diffusion imparted by boundaries or nonuniformities in the stress ratio.

\subsection{Interplay between hysteresis and nonlocality}
\label{sec:InterplayBetweenHysteresisNonlocality}

We now investigate qualitatively the combined effects of non-monotonicity and nonlocal diffusion on the characteristics of the flow-arrest transition in the presence of a spatially-varying stress ratio. To do so, we calculate quasi-steady, stress-driven predictions of the NGF model with hysteresis in the plane shear under gravity configuration pictured in Figure \ref{fig:Geometries}(b), where flow occurs along the $x$-direction and gravity acts orthogonally along the $z$-direction. A shear stress $\tau_\mathrm{w}$ and pressure $P_\mathrm{w}$ are applied at the top wall, imparting under quasi-steady conditions a constant shear stress $\tau(z) = \tau_\mathrm{w}$ and a nonuniform pressure $P(z) = P_\mathrm{w} + \phi \rho_\mathrm{s} G z$, where $\rho_\mathrm{s}$ is the grain density, $\phi$ the mean area packing fraction and $G$ the acceleration of gravity. The ratio of shear stress to pressure is thus given by
\begin{equation}
\mu(z) = \frac{\tau(z)}{P(z)} = \frac{\mu_\mathrm{w}}{1+z/\ell},
\label{eq:StressRatioPlaneGravity}
\end{equation}
where $\mu_\mathrm{w} = \tau_\mathrm{w}/P_\mathrm{w}$ is the applied stress ratio at the top wall, and $\ell = P_\mathrm{w}/\phi \rho_\mathrm{s} G$ is a loading length scale measuring the relative importance of the pressure imparted by the top wall versus that due to the weight of the grains. Critically, $\ell$ is inversely proportional with the degree of nonuniformity of the stress ratio \eqref{eq:StressRatioPlaneGravity} and, thus, the strength of nonlocal effects in this geometry \cite{pouliquen2009}. Results from previous DEM simulations \cite{liu2018} as well as our own (see Section \ref{sec:PlaneShearGravity}) have shown that these nonlocal effects induce the same flow-arrest transition features (F1--F3) that are observed in other geometries.

We compute quasi-steady, time-dependent solutions of the NGF model with hysteresis using the same arbitrary parameters as in the previous section. Because the dynamics are uniform in the streamwise $x$-direction, the fluidity equation \eqref{eq:HysNGF2} reduces to a one-dimensional PDE for $g(z,t)$, which is discretized following the procedure presented in Appendix \ref{app:NumericalDiscretization}. From there, the strain rate $\dot{\gamma}(z,t)$ and therefore the velocity profile $v(z,t)$ can be computed using the flow rule \eqref{eq:HysNGF1}. The fluidity equation is driven by the stress ratio \eqref{eq:StressRatioPlaneGravity}, for which we choose an arbitrary value $\ell = 100d$ small enough that the results are independent of the height of the domain\footnote{Previous DEM simulations\cite{liu2018} have shown that the vertical extent of the shear region below the top wall scales with $\ell$. For small enough $\ell$, the bottom region is thus quasi-static, and the results are independent of domain height.}. Following previous work \cite{liu2018}, the influence of boundaries is minimized by prescribing homogeneous Neumann boundary condition for $g$ at both walls. Simulations begin in a flowing state at $\mu_\mathrm{w} = 0.35$, then $\mu_\mathrm{w}$ is progressively decreased to 0.25 before being ramped back up to 0.35 in order to induce flow arrest and restart. We ensure that the ramp rate is slow enough that it does not affect the results. Most importantly, we perform these simulations for various values of the scalar parameter $A$ prescribing the strength of nonlocal effects, so that we can pinpoint the specific contributions of nonlocal diffusion and non-monotonicity of the limiting local rheology \eqref{eq:LocalRheologyNonmonotonic} to each of the three features (F1--F3).

\begin{figure*}[htb]
 \centering
 \includegraphics[width=\textwidth]{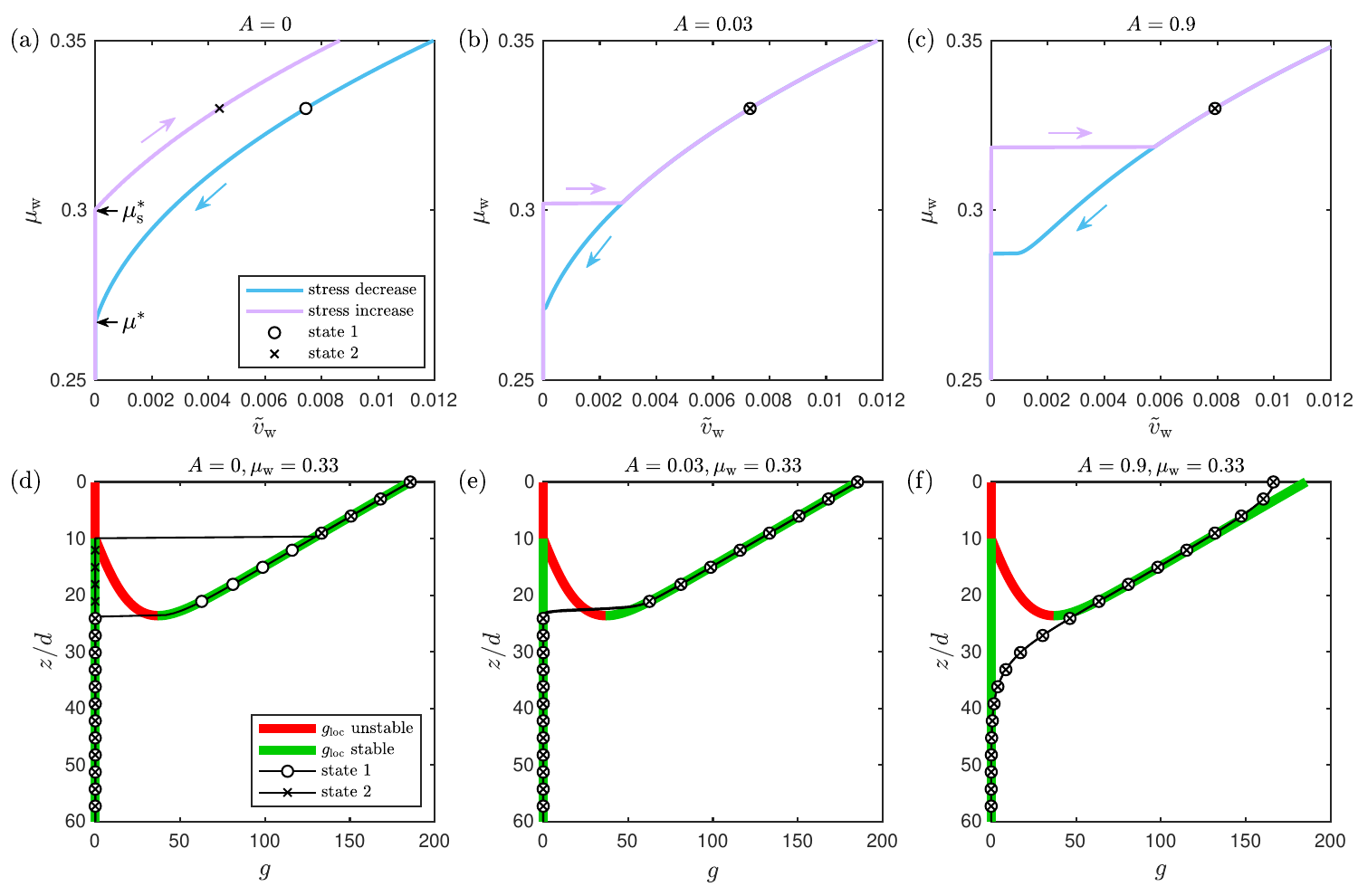}
 \caption{Qualitative behavior of the NGF model with hysteresis in stress-driven plane shear with gravity. (a,b,c) Time-dependent dimensionless velocity at the top wall, $\tilde{v}_\mathrm{w}(t)$, versus the applied stress ratio at the wall, $\mu_\mathrm{w}(t)$. (d,e,f) Fluidity field $g(z,t)$ corresponding to the state indicated in (a,b,c) by the lone circle (cross) on the down (up) branch at $\mu_\mathrm{w} = 0.33$. The field belonging to the down (up) branch is shown with circle (cross) markers and is superimposed to the stable (green) and unstable (red) steady-state solutions $g_\mathrm{loc}$ of the local fluidity equation \eqref{eq:HysNGF2_loc} under the stress ratio field \eqref{eq:StressRatioPlaneGravity}. {A movie version of this figure, which follows the state of the system as $\mu_\mathrm{w}(t)$ is progressively decreased and increased, is also included in the electronic supplementary information.}}
 \label{fig:Schema}
\end{figure*}

Figures \ref{fig:Schema}(a--c) display the time-dependent dimensionless velocity at the top wall, $\tilde{v}_\mathrm{w}(t)$, versus the applied stress ratio, $\mu_\mathrm{w}(t)$, for (a) $A = 0$, (b) $A = 0.03$, and (c) $A = 0.9$.
Here, the dimensionless velocity is defined as $\tilde{v}_\mathrm{w}(t) = v(z=0,t)/\ell \sqrt{m/P_w}$. The down stress ramp is shown in blue while the up stress ramp is shown in black, as depicted by the arrows. Further, Figures \ref{fig:Schema}(d--f) display both $g$ fields corresponding to the two states indicated by the lone circle and cross on the down and up ramps at $\mu_\mathrm{w} = 0.33$. Correspondingly, these fields are shown with circle or cross markers depending on the stress ramp that they belong to, and they are superimposed to the stable (green) and unstable (red) steady-state solutions $g_\mathrm{loc}$ of the local fluidity equation \eqref{eq:HysNGF2_loc} under the same stress ratio and pressure fields. {A movie version of Figure \ref{fig:Schema}, which follows the state of the system as the applied stress ratio $\mu_\mathrm{w}(t)$ is progressively decreased and increased, is also included in the electronic supplementary information.}

We begin with the case $A = 0$, for which nonlocal effects are turned off and hence the fluidity equation \eqref{eq:HysNGF2} is identical with its local limit \eqref{eq:HysNGF2_loc}. As shown in Figure \ref{fig:Schema}(a), the non-monotonicity of the limiting local rheology \eqref{eq:LocalRheologyNonmonotonic} leads to different $\mu_\mathrm{w}$ versus $v_\mathrm{w}$ branches in Figure \ref{fig:Schema}(a) when the stress is ramped down or up. Indeed, the bistable behavior of \eqref{eq:HysNGF2_loc} for $\mu^* < \mu < \mu_\mathrm{s}^*$ implies that there are two stable steady-state solutions $g_\mathrm{loc}$ -- one flowing and one arrested -- within a range of heights, as shown by the green branches in Figure \ref{fig:Schema}(d). When the applied stress is ramped down, the bistable region moves towards smaller {(shallower)} values of $z$, that were previously flowing; thus, the time-dependent solution {for $g$} will remain on the flowing branch. Conversely, when the applied stress is ramped up, the bistable region moves towards larger {(deeper)} values of $z$, that were previously arrested; thus, the time-dependent solution will remain on the arrested branch. This explains why the down ramp flows at a higher wall velocity $v_\mathrm{w}$ than the up ramp in Figure \ref{fig:Schema}(a), which also causes flow to arrest at a lower wall stress ratio, $\mu^*$, than that at which it restarts, $\mu_\mathrm{s}^*$. {Further, the onset of flow during the up ramp is characterized by a smooth increase in the top wall velocity since the thickness of the flowing layer beneath the top wall smoothly increases from zero (see movie).} In conclusion, the non-monotonicity of the local rheology suffices to endow the flow-arrest transition with feature (F1), but features (F2) and (F3) are absent -- there is no finite velocity jump at flow onset, and the starting and stopping stresses are identical with the local rheology predictions in Figure \ref{fig:LocalRheology}(c).

We then turn to the case $A = 0.03$, corresponding to a tiny amount of nonlocal effects. Figure \ref{fig:Schema}(b) suggests that in this situation, the main role of the nonlocal diffusion term in \eqref{eq:HysNGF2} is to merge the flowing section of the up stress ramp with that of the down stress ramp, which is almost unchanged from the case $A = 0$. In other words, there is only one possible flowing solution for every value of $\mu_\mathrm{w}$, and Figure \ref{fig:Schema}(e) shows that in the bistable range of heights, this unique solution follows the flowing local solution $g_\mathrm{loc}$ and not the arrested one. This is a direct consequence of the regularizing effect of the diffusion term, which acts to minimize discontinuities in the fluidity profile. A crucial side effect is that a finite velocity jump emerges at flow onset, since the entire bistable region beneath the top wall suddenly jumps from the arrested to the flowing local solution {(see movie)}. The interplay between nonlocality and non-monotonicity of the local rheology is therefore critical in achieving feature (F2), with (F3) the only one that remains unaccounted for.

Finally, we investigate the case $A = 0.9$, corresponding to the real calibrated value that we use later. Figure \ref{fig:Schema}(b) shows that similar to the case $A = 0.03$, there is only one possible flowing solution for every value of $\mu_\mathrm{w}$. However, the increased strength of nonlocal diffusion leads to a different $\mu_\mathrm{w}$ versus $v_\mathrm{w}$ relationship than before, with much higher wall stress ratios at flow arrest and onset. This is caused by the diffusion term spreading fluidity towards the  $\mu < \mu^*$ region where the local solution is arrested, as revealed in Figure \ref{fig:Schema}(f). Thus, when $\mu_\mathrm{w}$ is hardly higher than the stopping and starting stress ratios observed in the case $A = 0.03$, the $\mu < \mu^*$ region acts as a fluidity sink that prevents the overall nonlocal solution from flowing. The resulting strengthening of the wall stress ratio at flow onset and arrest is dependent on the degree of nonuniformity of the stress ratio \eqref{eq:StressRatioPlaneGravity} controlled by the loading length scale $\ell$, and increases with decreasing $\ell$. As a result, nonlocal diffusion induces strengthening of the threshold for flow onset and arrest with reducing system size, which generalizes a similar conclusion from previous studies \cite{kamrin2015,liu2017} that only looked at the stopping stress.

To summarize, non-monotonicity and nonlocality are seen to contribute in different ways to the features (F1--F3) of the flow arrest transition in the case of plane shear with gravity:
\begin{enumerate}[label=(F\arabic*),leftmargin=25pt,itemsep=-2pt]
\item hysteresis of the critical stresses for flow onset and arrest is achieved through non-monotonicity of the local rheology;
\item the finite velocity jump at flow onset requires an interplay between non-monotonicity and nonlocal diffusion;
\item increased strengthening at smaller system sizes is caused by nonlocal diffusion.
\end{enumerate}
Clearly, all three features are simultaneously achievable only when both non-monotonicity and nonlocality are included in the model. Even though these results have been obtained in a plane shear under gravity configuration, they should hold in other geometries that display a similar spatially-varying stress ratio profile such as annular shear between concentric cylinders, since the mechanisms at play are similar \cite{pouliquen2009,liu2018}. Finally, the conclusions that we have reached should apply to other nonlocal rheological models, including in particular those that treat the inertial number $I$ as an order parameter in place of $g$ \cite{bouzid2013,lee2017}.

\section{Comparisons with DEM simulations}

Now that we have established that the NGF model with hysteresis is able to reproduce qualitatively the various features of the flow-arrest transition, the next step is to compare quantitatively model predictions with discrete element method (DEM) simulations in a variety of geometries. To do so, we first calibrate the rheological parameters of the model using DEM simulations of homogeneous, simple plane shear. The calibrated model is then compared with DEM simulations in plane shear under gravity and inclined plane geometries.

We first describe the general setup of our DEM simulations, which we perform in the open-source software LAMMPS \cite{plimpton1995}. We simulate 2D disks with mean diameter $d = 0.0008 \, \mathrm{m}$ and aerial density $\rho_\mathrm{s} = 1.3 \, \mathrm{kg/m^2}$, corresponding to a characteristic grain mass $m = \rho_\mathrm{s} \pi d^2/4$. The disk diameters are uniformly distributed within $\pm 20 \%$ of $d$ in order to prevent crystallisation. Following seminal previous work \cite{dacruz2005}, we use the standard spring-dashpot model with Coulomb friction for the contact force between overlapping particles \cite{cundall1979}. More specifically, the normal force is given by $F_n = k_n \delta_n + g_n \dot{\delta}_n$ where $\delta_n \ge 0$ is the normal contact overlap, $k_n$ the normal stiffness and $g_n$ the damping coefficient, which can be expressed in terms of the coefficient of restitution $e$ for a binary collision as $g_n = - (mk_n)^{1/2} (2 \ln e)/(2(\pi^2+\ln^2 e))^{1/2}$. The tangential force is given by $F_t = k_t \delta_t$ where $\delta_t$ is the accumulated tangential contact displacement and $k_t$ is the tangential stiffness, and its magnitude is limited by the surface friction coefficient $\mu_\mathrm{surf}$ so that $|F_t| \le \mu_\mathrm{surf} |F_n|$. Thus, the contact force model is fully described by the parameters $k_n$, $k_t$, $e$, and $\mu_\mathrm{surf}$, to which we assign the same values as in past studies \cite{koval2009,kamrin2014,liu2018}. Namely, we use $\mu_\mathrm{surf} = 0.4$ and we choose $k_n$ so that $\kappa = k_n/P > 10^4$, with $P$ the characteristic confining pressure, corresponding to the stiff grain limit \cite{roux2002,dacruz2005}. Further, we set $k_t/k_n = 1/2$ and $e = 0.1$, with both having little influence on the phenomenology of dense flows of stiff disks \cite{silbert2001,campbell2002}. Finally, we choose a time step equal to 0.1 times the binary collision time $\tau_c = (m(\pi^2+\ln^2 e)/4k_n)^{1/2}$. At the end of each simulation, the particle-wise quantities in each saved system snapshot are coarse-grained into continuum fields according to the procedure described in Appendix \ref{app:CoarseGrainingProcedure}. Because the geometries that we investigate are homogeneous along the $x$-direction, this spatial averaging procedure returns an instantaneous streamwise velocity field $v(z,t)$, as well as instantaneous stress field components $\sigma_{xx}(z,t)$, $\sigma_{zz}(z,t)$, and $\sigma_{xz}(z,t)$.

\subsection{Calibration with simple plane shear}
\label{sec:CalibrationSimplePlaneShear}

We first simulate a simple plane shear geometry in DEM, since the homogeneity of the flow in this configuration enables us to calibrate the local part of the NGF model, given by the limiting local rheology \eqref{eq:LocalRheologyNonmonotonic}. The configuration is pictured in Figure \ref{fig:Geometries}(a), and consists of two parallel walls of length $L = 120d$ aligned along the horizontal $x$-direction, and separated by a distance $H$ along the vertical $z$-direction. The walls consist of glued disks, colored in brown in Figure \ref{fig:Geometries}(a), and they shear a dense collection of enclosed disks, colored according to their relative velocity magnitude in a particular snapshot of a flowing state. Periodic boundary conditions are applied along the $x$-direction, and the absence of gravity leads to a uniform stress ratio throughout, which is imparted by the walls. The top wall is assigned a horizontal velocity $v_\mathrm{w}$ that is either directly prescribed, corresponding to a velocity boundary condition, or calculated following a control scheme that simulates an applied tangential stress $\tau_\mathrm{w}$ to the top wall through the feedback law $\dot{v}_\mathrm{w} = (\tau_\mathrm{w} - \sigma_{xz}(z=0,t)) L/M_\mathrm{w}$, corresponding to a stress boundary condition. Here, the instantaneous tangential stress $\sigma_{xz}$ exerted by the flowing grains is directly evaluated at the wall, and the effective wall mass $M_\mathrm{w}$ acts as a damping parameter, which we take as $M_\mathrm{w} = 2000m$. Although velocity-driven DEM simulations of plane shear flow are the norm \cite{midi2004}, such simulations miss important features of the flow-arrest transition \cite{srivastava2019}. To our knowledge, stress-driven plane shear simulations have only been implemented in very few studies either through a solid wall \cite{volfson2003,ciamarra2011}, which corresponds to our setup, or through shearing of the periodic boundaries \cite{otsuki2011,smith2014,srivastava2017,srivastava2021b}. Finally, the pressure at the top wall is maintained close to a target value $P_\mathrm{w}$ through a widely used feedback law \cite{dacruz2005} according to which the distance $H$ between the walls evolves as $\dot{H} = (P_\mathrm{w} + \sigma_{zz}(z=0,t))L/g_\mathrm{w}$, where the instantaneous normal stress $\sigma_{zz}$ exerted by the flowing grains is directly evaluated at the wall, and $g_\mathrm{w}$ is a damping parameter that we take as $g_\mathrm{w} = 100 (mk_n)^{1/2}$.

\begin{figure*}[htb]
 \centering
 \includegraphics[width=\textwidth]{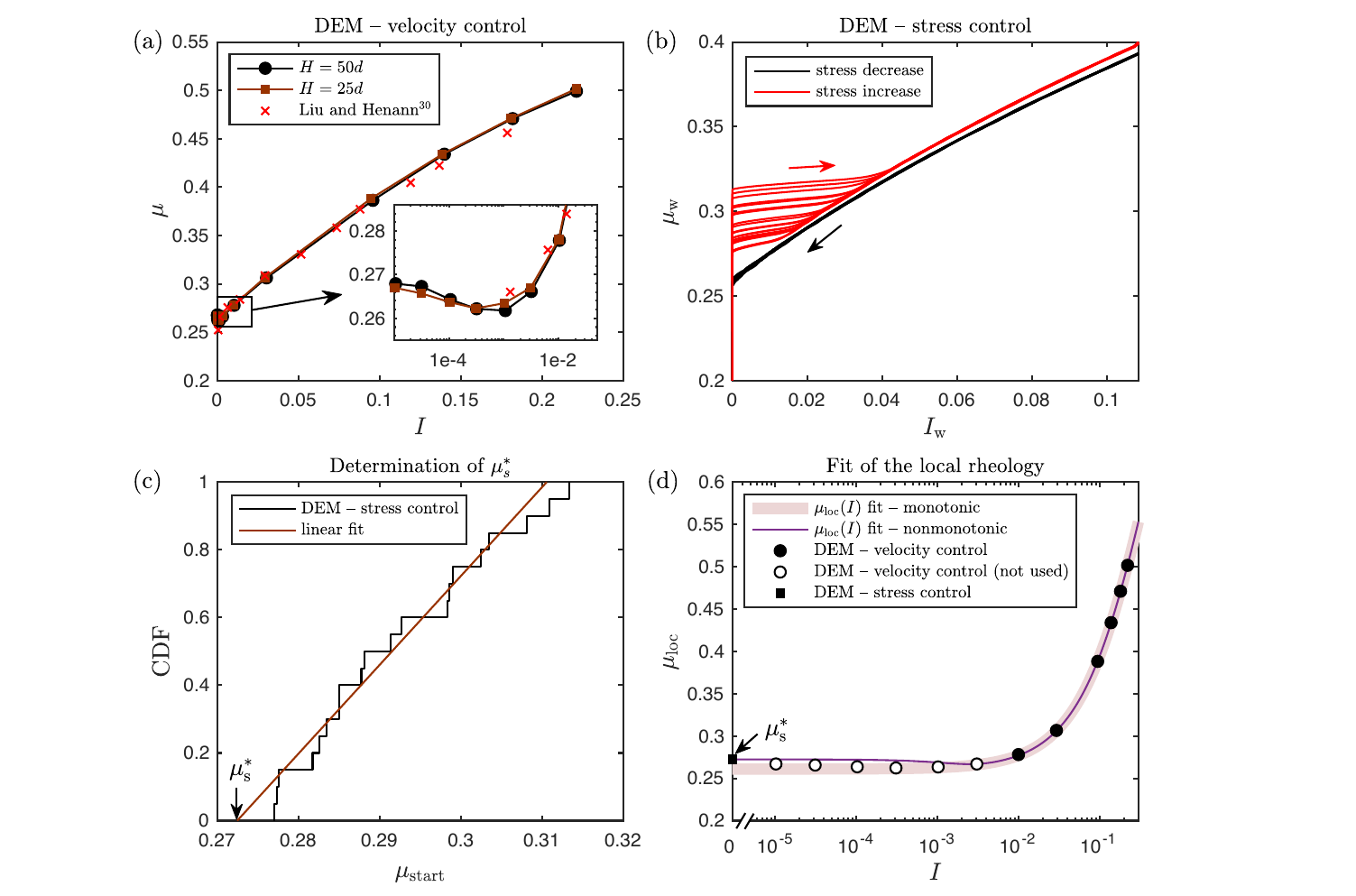}
 \caption{Calibration of the local part of the NGF model using simulations of plane shear without gravity. (a) Local $\mu(I)$ rheology obtained from velocity-driven DEM simulations for two system sizes, compared with data from Liu and Henann \cite{liu2018}. (b) Applied stress ratio at the top wall, $\mu_\mathrm{w}$, versus wall-based inertial number, $I_\mathrm{w}$, obtained from stress-driven DEM simulations for $H = 50d$, in which the stress ratio applied to the upper wall is first slowly ramped down (black lines), then ramped back up (red lines). Different lines correspond to different initial conditions, realized by letting the system flow at $\mu = 0.4$ for varying amounts of time. (c) Determination of the local yield stress ratio $\mu_\mathrm{s}^*$ from the stochastic $\mu_\mathrm{start}$ values pertaining to the different realizations. (d) The limiting non-monotonic local rheology \eqref{eq:LocalRheologyNonmonotonic} is fit in two steps. First, the parameters shared with the monotonic form \eqref{eq:LocalRheologyMonotonic} are calibrated using the velocity-driven DEM data for $I \ge 10^{-2}$ (filled circles), producing the monotonic $\mu_\mathrm{loc}(I)$ fit. Then, the parameters of the strain-rate weakening term $\chi(I;\kappa)$ are chosen so that $\mu_\mathrm{s}^*$ is equal to the value extracted from the stress-driven DEM data (filled square) and the minimum of $\mu$ occurs for $10^{-3} < I^* < 10^{-2}$, finally producing the non-monotonic $\mu_\mathrm{loc}(I)$ fit.}
 \label{fig:Calibration}
\end{figure*}

We begin by performing velocity-driven DEM simulations under various prescribed values of the top-wall velocity $v_\mathrm{w}$ and for two nominal system heights $H = 50d$ and $25d$. After each simulation has reached a steady state, we save 4000 system snapshots evenly distributed in time over a minimum additional top-wall shear displacement of 78$H$. The instantaneous continuum velocity and stress fields produced by the coarse-graining procedure are then averaged in time. As expected from numerous previous studies \cite{midi2004,dacruz2005}, the strain rate and the stress components are all approximately constant in the central part of the sheared layer, four grain diameters away from the walls. We therefore spatially average these quantities into the strain rate $\dot{\gamma} = \langle |dv/dz| \rangle$, shear stress $\tau = \langle \sigma_{xz} \rangle$ and pressure $P = \langle | \sigma_{zz} | \rangle$, from which we calculate the stress ratio $\mu = \tau/P$ and the inertial number $I = \dot{\gamma} \sqrt{m/P}$. Different values of the prescribed top-wall velocity produce different $(\mu,I)$ pairs, which are displayed in Figure \ref{fig:Calibration}(a) for the two nominal system heights considered.
We also plot corresponding results from the DEM simulations of Liu and Henann \cite{liu2018}, obtained under identical system parameters and for a nominal height $H = 60d$. The agreement between the two sets of data validates our simulations; furthermore, the negligible difference between the $\mu(I)$ curves pertaining to different system heights demonstrates the negligible influence of the walls. Importantly, we observe the presence of a strain-rate weakening regime at low enough values of $I$, directly corroborating previous simulation results from DeGiuli and Wyart \cite{degiuli2017}. However, a decreasing relationship between shear stress and strain rate is mechanically unstable and typically results in the formation of shear bands that, in turn, render impossible the accurate measurement of the true strain rate in the strain-rate weakening regime \cite{divoux2016}. Furthermore, past theoretical studies \cite{yuan1999,lu2000} suggest that in this regime, systems with a nonlocal flow rule select a specific stress state independent of the nominal strain rate imparted by the walls. The NGF model, therefore, cannot be simply calibrated on velocity-driven DEM data if it is to accurately predict onset and arrest of flow under variations of the applied stress. 

In order to extract the true critical stresses delineating the transition between static and flowing regimes, we run stress-driven DEM simulations of flow onset and arrest under slowly decreasing and increasing ramps of the top-wall stress ratio $\mu_\mathrm{w}$, for a nominal system height $H = 50d$. Specifically, we assign a time-dependent top-wall shear stress $\tau_\mathrm{w}(t) = \mu_\mathrm{w}(t) P_\mathrm{w}$ according to the control procedure described above, where $\mu_\mathrm{w}(t)$ is the target stress ratio applied to the top wall and $P_\mathrm{w}$ is the constant target pressure. We run 20 different simulations by letting the system flow at $\mu_\mathrm{w} = 0.4$ for a varying amount of time after it has reached steady state, effectively imparting a different initial microstructure to each simulation. The applied stress ratio $\mu_\mathrm{w}$ is then linearly decreased from 0.4 to 0.25 over a time duration of $6.5 \cdot 10^7 \tau_c \simeq 113 \, \mathrm{s}$, inducing jamming of the grains. We then let the contact forces relax by decreasing $\mu_\mathrm{w}$ from 0.25 to 0 over a time duration of $2 \cdot 10^7 \tau_c \simeq 35 \, \mathrm{s}$ and keeping $\mu_\mathrm{w}$ at 0 for $1 \cdot 10^7 \tau_c \simeq 17 \, \mathrm{s}$. Finally, $\mu_\mathrm{w}$ is linearly ramped back up, first from 0 to 0.25 over a time duration of $2 \cdot 10^7 \tau_c \simeq 35 \, \mathrm{s}$, then from 0.25 to 0.4 over $6.5 \cdot 10^7 \tau_c \simeq 113 \, \mathrm{s}$, triggering onset of flow. The ramp rate is slow enough that the system can be assumed to undergo quasi-steady motion. We save 4000 system snapshots evenly distributed in time, from start to end of the stress ramp. At the end of each simulation, we calculate an instantaneous wall-based inertial number $I_\mathrm{w}(t) = v_\mathrm{w}(t)/H_\mathrm{w} \sqrt{m/P_\mathrm{w}}$, where $v_\mathrm{w}(t)$ is a moving time window average over 50 snapshots of the instantaneous wall velocity to smooth out small fluctuations, and $H_\mathrm{w}$ is the average true vertical distance between both walls. Thanks to the little amount of observed wall slip\footnote{In our DEM simulations, the relative difference between the upper wall velocity and the coarse-grained streamwise velocity of the grains near the upper wall never exceeds 3\%.} and the quasi-steady conditions, we may consider both $\mu_\mathrm{w}$ and $I_\mathrm{w}$ as smooth approximations of the highly-fluctuating instantaneous values of $\mu$ and $I$ in the bulk. The resulting $\mu_\mathrm{w}$ versus $I_\mathrm{w}$ curves are shown in Figure \ref{fig:Calibration}(b) in black and red for the decreasing and increasing stress ramps, respectively, with different curves corresponding to different simulations. Observe the similarity between these curves and the ones shown in Figure \ref{fig:Geometries}(a), with hysteresis (feature F1) and a velocity jump at flow onset (feature F2) clearly visible in both geometries. Besides, our simulations reveal that the critical stress ratio at flow onset is stochastic and noticeably higher than the threshold obtained from velocity-driven simulations in Figure \ref{fig:Calibration}(a) as $I$ vanishes. Such variability in the transition between arrested and flowing states has been observed previously \cite{clark2015}, and may be explained by the role played by the specific structure of the particle contact network \cite{bi2011,clark2018,srivastava2021}. On the other hand, the critical stress ratio at flow arrest is much more narrowly distributed and similar to the velocity-driven flow threshold.

Thus, the NGF model needs to be calibrated using data from both velocity-driven and stress-driven DEM simulations, so as to correctly predict the characteristics of both the flowing regime and the transition between arrested and flowing states. However, the critical stress ratio for flow onset observed in the stress-driven simulations is highly stochastic, while the limiting local rheology \eqref{eq:LocalRheologyNonmonotonic} of the model predicts a deterministic value $\mu_\mathrm{s}^*$. {To reconcile this apparent contradiction, we note that because continuum models in general aim to reproduce the ensemble-average behavior of the discrete system across all possible realizations, we expect our NGF model to predict onset of flow so long as any measurable region of the space of ensembles initiates flow. With this in mind,} we therefore reduce the stochastic critical stress from DEM to a unique deterministic value that represents the \textit{lowest achievable} starting stress of the system as follows. For each realization, we first define $\mu_\mathrm{start}$ as the observed $\mu_\mathrm{w}$ when $I_\mathrm{w}$ last exceeds $10^{-3}$ during stress increase. We then assume that the ensemble of stochastic $\mu_\mathrm{start}$ values follows a uniform probability distribution over a finite range (bounded from below by the lowest achievable starting stress), which allows us to fit a linear function to their cumulative distribution (CDF), shown in Figure \ref{fig:Calibration}(c). An estimate for the lowest achievable $\mu_\mathrm{start}$ is given by the $x$-intercept of the fitted CDF, which we thus assign to $\mu_\mathrm{s}^* = 0.2724$. Finally, we show in Appendix \ref{app:SizeEffectsSimplePlaneShear} that the distribution of stochastic $\mu_\mathrm{start}$ values barely changes for smaller nominal heights $H = 25d$ and $10d$, which supports our methodology of calculating the local yield stress ratio $\mu_\mathrm{s}^*$ predicted by the model using stress-driven simulations at $H = 50d$.

The parameters of the limiting local rheology \eqref{eq:LocalRheologyNonmonotonic} can now be calibrated following a two-step approach pictured in Figure \ref{fig:Calibration}(d). First, the parameters shared with the monotonic form \eqref{eq:LocalRheologyMonotonic} are calibrated using the velocity-driven DEM data corresponding to $H = 50d$ and $I \ge 10^{-2}$ (filled circles), producing the monotonic $\mu_\mathrm{loc}(I)$ fit. The velocity-driven DEM data for $I < 10^{-2}$ (open circles) is also shown for reference, but is not used in the calibration. Second, the parameters of the strain-rate weakening term $\chi(I;\kappa)$ are chosen so that $\mu_\mathrm{s}^*$ is equal to the value extracted from the stress-driven DEM data (filled square) and the minimum of $\mu$ occurs for $10^{-3} < I^* < 10^{-2}$, finally producing the non-monotonic $\mu_\mathrm{loc}(I)$ fit. The resulting parameter values are $\mu_\mathrm{s} = 0.2610$, $\mu_2 = 0.9784$, $b = 1.6406$, $a = 0.0116$, $c = 50$, and $n = 1/4$. We note that once $\mu_\mathrm{s}$ and $\mu_2$ are known from the first step, $a$ is obtained in the second step from $\mu_\mathrm{s}^*$ by inverting \eqref{eq:YieldStressNonmonotonic}. In the second step, we have selected the value $1/4$ from DeGiuli and Wyart \cite{degiuli2017} for the parameter $n$ that controls the grain stiffness-dependence of the hysteresis amplitude. We have nonetheless verified that choosing instead $n = 0$, which removes the stiffness dependence, and recalibrating $c$ accordingly produces negligible changes in our results to follow. Finally, we adopt the value $A = 0.9$ from Liu and Henann \cite{liu2018} for the nonlocal amplitude, which they calibrated on DEM data obtained with the same particle and contact force law properties.

\subsection{Plane shear under gravity}
\label{sec:PlaneShearGravity}

We now compare the predictions of the calibrated NGF model with stress-driven DEM simulations of plane shear under gravity shown in Figure \ref{fig:Geometries}(b). We have already investigated in Section \ref{sec:InterplayBetweenHysteresisNonlocality} the qualitative behavior of the model in this geometry, in which the gravity field imparts a nonuniform distribution of stress ratio $\mu(z)$ characterized by a loading length scale $\ell$; see equation \eqref{eq:StressRatioPlaneGravity}. Besides the presence of gravity, the DEM setup of the system is identical to that of the previous section, except for the nominal distance $H = 60d$ between the walls. The shear stress and pressure at the top wall are controlled according to the feedback schemes described in the previous section, and the bottom wall is fixed.

Similar to the case of simple plane shear, we perform DEM simulations of arrest and onset of flow under decreasing then increasing ramps of applied stress. Specifically, a time-varying top-wall stress ratio $\mu_\mathrm{w}(t)$ is prescribed through a time-dependent target shear stress $\tau_w(t) = \mu_\mathrm{w}(t) P_\mathrm{w}$ and constant target pressure $P_w$. The applied stress ratio $\mu_\mathrm{w}(t)$ follows the same protocol as in Section \ref{sec:CalibrationSimplePlaneShear}, with the only difference being that we start from, and end at, a top value of $\mu_\mathrm{w} = 0.45$ instead of $\mu_\mathrm{w} = 0.4$. Correspondingly, the duration of the decreasing and increasing sections of the stress ramp between $0.45$ and $0.25$ is lengthened to $9 \cdot 10^7 \tau_c \simeq 157 \, \mathrm{s}$ so that the rate of change of $\mu_\mathrm{w}(t)$ is unaffected. As before, for each length scale $\ell$ we run 20 different simulations corresponding to different initial microstructures, by letting the system spend a varying amount of time in the initial shearing period at $\mu_\mathrm{w} = 0.45$. During that period, gravity is first turned off to ensure homogeneous mixing of the grains, before being turned back on. We save 4000 system snapshots for each simulation, from which we calculate $v_\mathrm{w}(t)$, a moving time window average over 50 snapshots of the instantaneous wall velocity. We then calculate the nondimensional wall velocity as $\tilde{v}_\mathrm{w}(t) = v_\mathrm{w}(t)/\ell \sqrt{m/P_\mathrm{w}}$. In order to compare deterministic predictions from the NGF model with the DEM results corresponding to all $N = 20$ different runs, we transform the discrete values $\tilde{v}_\mathrm{w}^{(i)}(t)$ for $i = 1, \dots, N$ at each time step into a continuous probability density function (PDF) as follows:
\begin{equation}
f(\tilde{v}_\mathrm{w}) = \int_{-\infty}^\infty w(v - \tilde{v}_\mathrm{w}) \left[ \frac{1}{N} \sum_{i = 1}^{N} \delta(v - \tilde{v}_\mathrm{w}^{(i)}) \right] dv,
\label{eq:PDF}
\end{equation}
where $\delta$ is the Dirac delta function, and $w$ is the Gaussian kernel
\begin{equation}
w(v) = \frac{1}{\sqrt{2\pi} L} e^{-x^2/2L^2},
\end{equation}
with $L$ the kernel size, which we choose equal to 0.01 times the maximum observed value of $\tilde{v}_\mathrm{w}^{(i)}(t)$ across all runs and times. The PDF \eqref{eq:PDF} can be conveniently expressed as
\begin{equation}
f(\tilde{v}_\mathrm{w}) = \frac{1}{N} \sum_{i=1}^{N} w(\tilde{v}_\mathrm{w} - \tilde{v}_\mathrm{w}^{(i)}),
\end{equation}
and it integrates to one, as it should. Further, we also need to reduce the stochastic transition stresses between flowing and arrested states into a unique deterministic value, which we define in a similar way to Section \ref{sec:CalibrationSimplePlaneShear} as the \textit{lowest achievable} value to be consistent with the methodology followed in calibrating the local yield stress ratio $\mu_\mathrm{s}^*$ of the model. For each realization, we first calculate $\mu_\mathrm{w,start}$ as the observed $\mu_\mathrm{w}$ when $\tilde{v}_\mathrm{w}$ last exceeds the threshold value $10^{-3}$ during stress increase, and $\mu_\mathrm{w,stop}$ as the observed $\mu_\mathrm{w}$ when $\tilde{v}_\mathrm{w}$ last falls below the same threshold during stress decrease. The deterministic starting and stopping critical stress ratios are then defined as the $x$-intercept of a fitted linear CDF to the stochastic $\mu_\mathrm{w,start}$ and $\mu_\mathrm{w,stop}$ values, similar to Figure \ref{fig:Calibration}(c). In the following, we will refer to these deterministic thresholds as $\mu_\mathrm{w,start}$ and $\mu_\mathrm{w,stop}$.

We also compute NGF model predictions in the same geometry, using the calibrated parameter values from Section \ref{sec:CalibrationSimplePlaneShear}. As described in Section \ref{sec:InterplayBetweenHysteresisNonlocality}, the fluidity equation \eqref{eq:HysNGF2} reduces to a one-dimensional PDE for $g(z,t)$ that requires the stress ratio profile $\mu(z,t)$ as input. Thanks to the homogeneous and quasi-steady conditions, the latter is given through \eqref{eq:StressRatioPlaneGravity} and set by the top-wall stress ratio $\mu_\mathrm{w}(t)$, to which we assign the exact same temporal protocol as in the DEM simulations. Since the NGF model is not expected to be valid in the vicinity of the walls, we end the corresponding simulation domain at a distance $d_\mathrm{w} = 2d$ away from the real walls. Furthermore, we follow previous work \cite{liu2018} in using homogeneous Neumann ($\partial g / \partial z = 0$) boundary conditions on $g$ at the walls in order to minimize their influence.  The details of the discretization method for the ODE governing $g(z,t)$ are presented in Appendix \ref{app:NumericalDiscretization}. Once $g(z,t)$ is known, the strain rate $\dot{\gamma}(z,t)$ can be calculated using the flow rule \eqref{eq:HysNGF1} and integrated into the velocity profile $v(z,t)$, taking into account a slip length equal to $d_\mathrm{w}$ for the velocity at the bottom wall. Finally, the top-wall velocity is extrapolated as $v_\mathrm{w}(t) = v(z=-d_\mathrm{w},t) + d_\mathrm{w} \dot{\gamma}(z=-d_\mathrm{w},t)$ and is nondimensionalized into $\tilde{v}_\mathrm{w}(t)$.

\begin{figure*}[htb]
 \centering
 \includegraphics[width=\textwidth]{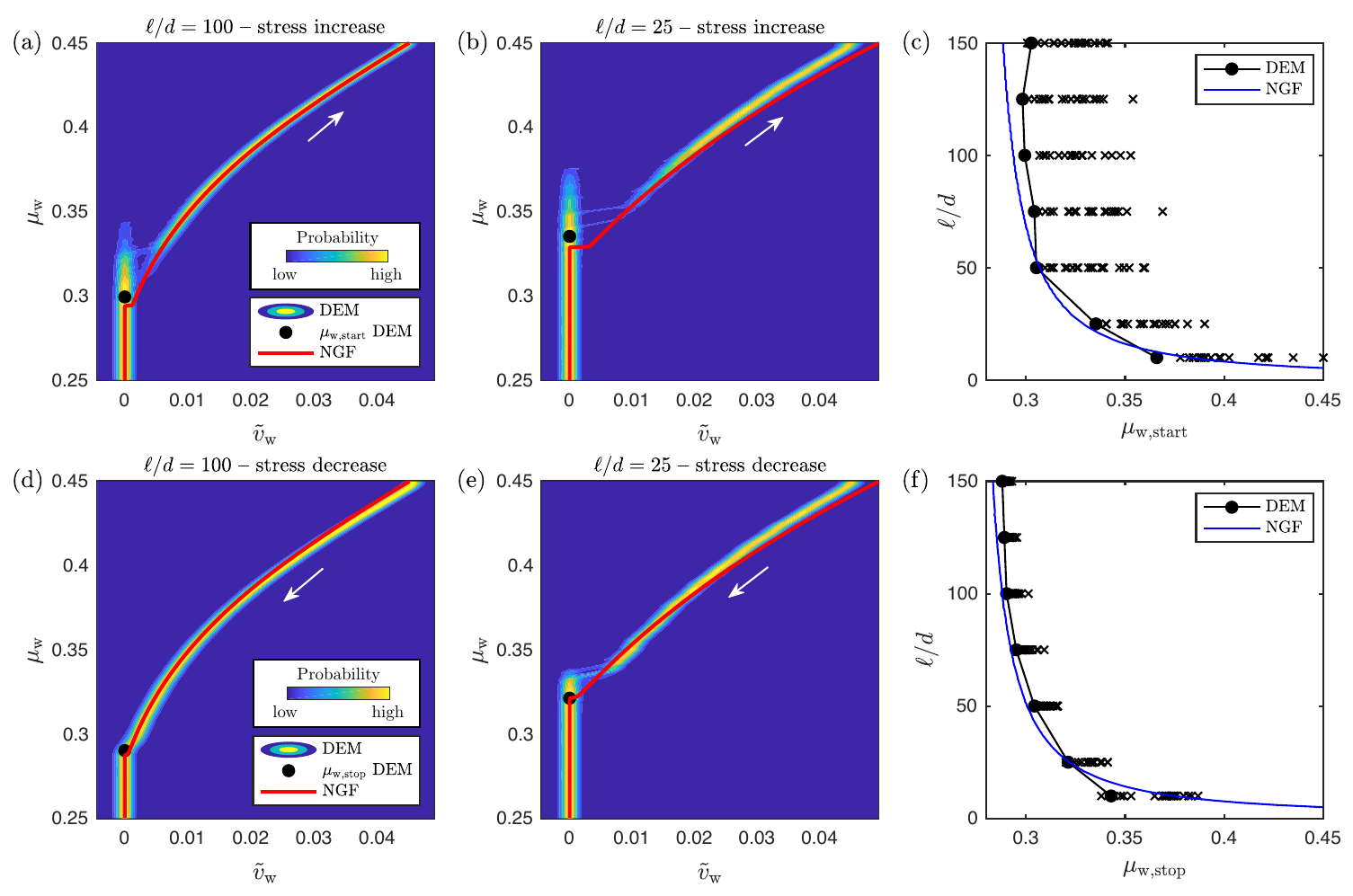}
 \caption{Comparison between NGF model predictions and DEM simulations for plane shear under gravity. (a,b,d,e) Relationship between stress ratio $\mu_\mathrm{w}$ and dimensionless velocity $\tilde{v}_\mathrm{w}$ at top wall from DEM (contour of probability values extracted from 20 different runs) and NGF (red lines) for two different loading length scales of (a,d) $\ell = 100d$ and (b,e) $\ell = 25d$, with the (a,b) increasing and (d,e) decreasing stress ramps shown separately. The filled circles display the deterministic $\mu_\mathrm{w,start}$ and $\mu_\mathrm{w,stop}$ critical stress ratios obtained from all DEM runs. (c,f) Critical stress ratios $\mu_\mathrm{w,start}$ and $\mu_\mathrm{w,stop}$ versus dimensionless loading length scale $\ell/d$ from DEM (filled circles) and NGF (continuous line). The crosses show the individual transition stresses pertaining to each of the 20 different runs, while the filled circles represent the deterministic value extracted from the linear fit of the CDF.}
 \label{fig:PlaneShearGravity}
\end{figure*}

Figures \ref{fig:PlaneShearGravity}(a,b,d,e) display the relationship between $\mu_\mathrm{w}$ and $\tilde{v}_\mathrm{w}$ obtained from both DEM simulations and NGF model predictions for two different loading length scales of (a,d) $\ell = 100d$ and (b,e) $\ell = 25d$.
The increasing stress ramp is shown in (a,b) while the decreasing stress ramp is shown in (d,e).  The DEM results are displayed as contours of $f(\tilde{v}_\mathrm{w})$ corresponding to each value of $\mu_\mathrm{w}$, in such a way that the plots can be read as the probability of occurrence of individual realizations, with yellow color indicating high probability and blue color indicating low probability. The deterministic $\mu_\mathrm{w,start}$ and $\mu_\mathrm{w,stop}$ values from DEM are also displayed as filled circles, and the NGF model prediction is shown as the red line. An excellent agreement between NGF and DEM is observed in the flowing regime\footnote{The small discrepancy observed in the case $\ell/d = 25$ is probably attributable to the choice of boundary conditions for $g$. An extensive discussion on the role of the latter is presented in Section \ref{sec:InclinedPlane}.}. Similarly, the transition between arrested and flowing states occurs at similar stress levels in both cases, and displays all three features (F1--F3) identified in the introduction: hysteresis between onset and arrest, velocity jump at onset, and strengthening with smaller $\ell/d$. The amount of velocity jump at flow onset exhibited by the NGF solution appears smaller than that observed in the DEM simulations; this is a consequence of the NGF model being calibrated so as to start flowing at the lowest possible critical stress based on the DEM simulations. 

Finally, Figures \ref{fig:PlaneShearGravity}(c,f) display the critical stresses $\mu_\mathrm{w,start}$ and $\mu_\mathrm{w,stop}$ versus the dimensionless loading length scale $\ell/d$. Shown are the individual transition stresses from all 20 DEM runs (crosses), the deterministic values extracted from the linear fit of the CDF (filled circles), and the corresponding NGF predictions (continuous line). The NGF predictions of the critical stresses are obtained using the methodology described in Appendix \ref{app:CriticalStressesNGF}, which circumvents the need to run time-dependent simulations for every value of $\ell/d$. Overall, the NGF model predicts a similar amount of strengthening as apparent in the DEM simulations. The slightly higher critical stresses exhibited in DEM for large loading length scales is caused by an observed change in the slope of the $g$ field at the boundary as the pressure applied by the top wall increases. Accordingly, implementing a homogeneous Robin boundary condition for $g$ with a finite associated length scale would lead the NGF model to predict higher values for the transition stresses, closing the gap with the DEM data. Lastly, Appendix \ref{app:HysteresisSizeDependence} shows that the hysteresis amplitude $\mu_\mathrm{w,start}-\mu_\mathrm{w,stop}$ is only weakly dependent on the dimensionless loading length scale $\ell/d$.

\subsection{Inclined plane}
\label{sec:InclinedPlane}
As a last example, we evaluate predictions from the calibrated NGF model against DEM simulations in the inclined plane configuration shown in Figure \ref{fig:Geometries}(c). A fixed basal wall of length $L = 120d$ and consisting of glued disks is inclined at an angle $\theta$ with respect to the horizontal, and is covered by a dense collection of freely moving disks forming a layer of height $H$. The $x$- and $z$-directions are parallel and orthogonal to the base wall, respectively, and periodic boundary conditions are applied along the $x$-direction. The gravity field imparts a uniform ratio of shear stress to pressure throughout the layer at steady state equal to $\mu = \tan \theta$, making this configuration inherently stress-driven. Even though $\mu$ is uniform, nonlocal effects still arise from the presence of the rough base, which acts as a sink for velocity fluctuations within the moving grains. The transition between onset and arrest of flow on an inclined plane exhibits all three features mentioned in the introduction, as documented in many past experimental and computational studies \cite{pouliquen1999,daerr1999,silbert2001,pouliquen2002,rognon2002,silbert2003,russell2019}. 

Following the previous cases, we run DEM simulations of flow arrest and onset by slowly decreasing then increasing the inclination angle. Specifically, we prescribe a temporal profile for $\theta(t)$ such that the stress ratio $\mu(t) = \tan \theta(t)$ follows the same protocol as in Section \ref{sec:CalibrationSimplePlaneShear}, with an initial flowing period at $\mu = 0.4$ followed by a continuous decrease to $\mu = 0$ and a continuous increase back to $\mu = 0.4$. As before, we execute 20 runs for each layer height $H$, each with a different time duration spent in the initial flowing regime at $\mu = 0.4$, giving a unique microstructure to every simulation before the start of the stress ramp. We save 4000 system snapshots in each simulation, from which we compute the instantaneous continuum velocity field $v(z,t)$. Anticipating that the NGF model will be run over a truncated domain ending at a distance $d_w = 2d$ away from the bottom wall and $d_s = 3d$ away from the layer's free surface, we then calculate a depth-averaged instantaneous velocity $\bar{v}(t)$ over the corresponding truncated domain, which we smooth out using a moving time window average over 50 snapshots. We then express $\bar{v}(t)$ in terms of a dimensionless Froude number defined as $\mathit{Fr}(t) = \bar{v}(t)/\sqrt{GH}$. As was done in the previous section, the discrete values $\mathit{Fr}^{(i)}(t)$ at each time step, for $i = 1, \dots, N$ corresponding to all $N = 20$ different runs, are transformed into a continuous PDF $f(\mathit{Fr})$ through the convolution \eqref{eq:PDF}. Finally, the stochastic transition stresses between flowing and arrested states are reduced into deterministic numbers $\mu_\mathrm{start}$ and $\mu_\mathrm{stop}$ according to the procedure detailed in Sections \ref{sec:CalibrationSimplePlaneShear} and \ref{sec:PlaneShearGravity}. For each realization, we first calculate $\mu_\mathrm{start}$ as the observed $\mu$ when $\mathit{Fr}$ last exceeds the threshold value $10^{-2}$ during stress increase, and $\mu_\mathrm{stop}$ as the observed $\mu$ when $\mathit{Fr}$ last falls below the same threshold during stress decrease. The deterministic starting and stopping critical stress ratios are then defined as the $x$-intercept of a fitted linear CDF to the stochastic $\mu_\mathrm{start}$ and $\mu_\mathrm{stop}$ values, similar to Figure \ref{fig:Calibration}(c). In the following, we will refer to these deterministic thresholds as $\mu_\mathrm{start}$ and $\mu_\mathrm{stop}$.

We compute predictions of the NGF model in the same geometry, using the calibrated parameter values from Section \ref{sec:CalibrationSimplePlaneShear} and the same temporal protocol for $\theta(t)$ as in the DEM simulations. The fluidity equation \eqref{eq:HysNGF2} reduces to a one-dimensional PDE for $g(z,t)$ that takes as input the stress ratio profile, which is still related to the inclination angle as $\mu(t) = \tan \theta(t)$ thanks to the quasi-steady conditions. The uniformity of the stress ratio implies that nonlocal effects are entirely imparted by boundaries, making the choice of boundary conditions critical. Similarly to our approach in Section \ref{sec:PlaneShearGravity}, the domain for the NGF solution is defined to start at a distance $d_\mathrm{w} = 2d$ away from the bottom wall due to the lack of validity of the NGF model near the boundary. At that location, the DEM data suggests a Robin-type homogeneous boundary condition for $g$, with an associated length scale $\delta$ that may be sensitive to various factors. We choose to sidestep the exact modeling of the boundary condition by considering the two edge cases of $\delta = 0$ and $\delta = \infty$, corresponding to homogeneous Dirichlet ($g = 0$) and Neumann ($\partial g / \partial z = 0$) boundary conditions, respectively. Regarding the top boundary, the DEM data shows that the strain rate vanishes about 3 grain diameters below the surface, corroborating previous studies \cite{silbert2001,silbert2003}. We thus end the NGF simulation domain at a distance $d_s = 3d$ away from the layer's free surface, and we prescribe a homogeneous Neumann ($\partial g / \partial z = 0$) boundary condition for $g$ there, following Kamrin and Henann \cite{kamrin2015}. We also assign a finite pressure to the top boundary equal to the weight of the neglected layer of thickness $d_s$, which is approximately equal to $P(z=-d_s) = 0.8 \rho_s G (2d) \cos \theta(t)$ due to the drop in packing fraction near the layer's surface. The details of the discretization method for the ODE governing $g(z,t)$ are presented in Appendix \ref{app:NumericalDiscretization}. Once $g(z,t)$ is known, the strain rate $\dot{\gamma}(z,t)$ can be computed through the flow rule \eqref{eq:HysNGF1}. Finally, the velocity can be integrated from $\dot{\gamma}(z,t)$, taking into account a velocity slip length equal to $d_\mathrm{w}$ at the bottom boundary of the NGF domain, and depth-averaged to produce $\bar{v}(t)$ and hence $\mathit{Fr}(t)$.

\begin{figure*}[htb]
 \centering
 \includegraphics[width=\textwidth]{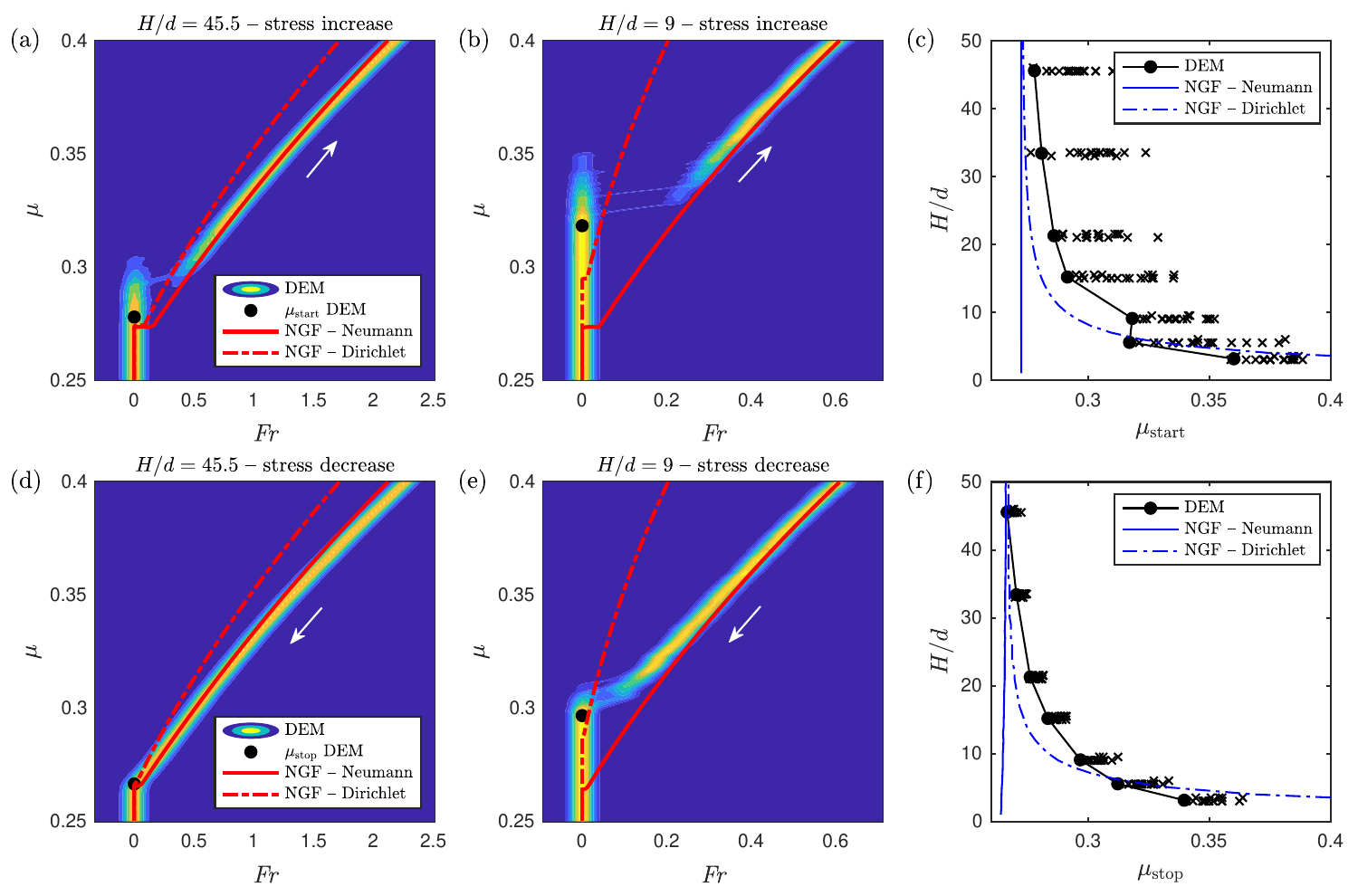}
 \caption{Comparison between NGF model predictions and DEM simulations for inclined plane. (a,b,d,e) Relationship between stress ratio $\mu = \tan \theta$ and Froude number $\mathit{Fr}$ from DEM (contour of probability values extracted from 20 different runs) and NGF (red lines) for two different layer heights of (a,d) $H = 45.5d$ and (b,e) $H = 9d$, with the (a,b) increasing and (d,e) decreasing stress ramps shown separately.  The filled circles display the deterministic $\mu_\mathrm{start}$ and $\mu_\mathrm{stop}$ critical stress ratios obtained from all DEM runs. (c,f) Critical stress ratios $\mu_\mathrm{start}$ and $\mu_\mathrm{stop}$ versus dimensionless layer height $H/d$ from DEM (filled circles) and NGF (blue lines). The crosses show the individual transition stresses pertaining to each of the 20 different runs, while the filled circles represent the deterministic value extracted from the linear fit of the CDF.}
 \label{fig:InclinedPlane}
\end{figure*}

Figures \ref{fig:InclinedPlane}(a,b,d,e) display the relationship between $\mu$ and $\mathit{Fr}$ obtained from both DEM simulations and NGF model predictions for two different layer heights at rest of (a,d) $H = 45.5d$ and (b,e) $H = 9d$.
The increasing stress ramp is shown in (a,b) while the decreasing stress ramp is shown in (d,e).  The DEM results are displayed as contours of $f(\mathit{Fr})$ corresponding to each value of $\mu$, in such a way that the plots can be read as the probability of occurrence of individual realizations, with yellow color indicating high probability and blue color indicating low probability. The deterministic $\mu_\mathrm{start}$ and $\mu_\mathrm{stop}$ values from DEM are also displayed as filled circles, and the NGF model predictions pertaining to the two basal boundary conditions for $g$ are shown as the continuous (homogeneous Neumann) and dash-dotted (homogeneous Dirichlet) red lines. The Neumann boundary condition in the NGF model leads to an excellent agreement with DEM in the flowing regime. However, it does not predict any strengthening of the critical transition stresses $\mu_\mathrm{start}$ and $\mu_\mathrm{stop}$ for smaller layer height $H$. The Dirichlet boundary condition, on the other hand, reproduces the strengthening of the transition stresses but fails to match the DEM results in the flowing regime.

The critical stresses $\mu_\mathrm{start}$ and $\mu_\mathrm{stop}$ are depicted in greater detail in Figures \ref{fig:InclinedPlane}(c,f) versus the dimensionless layer height $H/d$. Shown are the individual transition stresses from all 20 DEM runs (crosses), the deterministic values extracted from the linear fit of the CDF (filled circles), and the corresponding NGF predictions using homogeneous Neumann (continuous line) or homogeneous Dirichlet (dash-dotted line) boundary conditions. As we did for planar shear with gravity, the NGF predictions of the critical stresses are obtained using the methodology described in Appendix \ref{app:CriticalStressesNGF}, which bypasses the need to run time-dependent simulations for every value of $H/d$. The Neumann boundary condition does not produce any strengthening of the critical stresses, since it kills the principal cause of flow inhomogeneity in this geometry\footnote{In the presence of the Neumann boundary condition, a small amount of flow inhomogeneity is still incurred by the pressure-dependent quadratic term in the fluidity equation \eqref{eq:HysNGF2}, which explains the slight dependence of $\mu_\mathrm{stop}$ on $H/d$ in Figure \ref{fig:InclinedPlane}(f).}. On the other hand, the Dirichlet boundary condition generates a level of strengthening roughly comparable with the DEM data. As was the case for plane shear under gravity, Appendix \ref{app:HysteresisSizeDependence} shows that the hysteresis amplitude $\mu_\mathrm{w,start}-\mu_\mathrm{w,stop}$ is only weakly dependent on the dimensionless layer height $H/d$.

In summary, it appears that the critical stresses are best predicted by the NGF model endowed with a homogeneous Dirichlet boundary condition for $g$, while the flowing regime is most accurate when a homogeneous Neumann boundary condition is used. This dichotomy could stem either from missing ingredients in the NGF model itself or from an inaccurate description of the boundary condition. Regarding the former, a known shortcoming of the current formulation of the model is that it does not produce the widely documented collapse of the Froude number for all layer heights and angles \cite{pouliquen1999,silbert2003,forterre2003}. We have thus modified the fluidity equation \eqref{eq:HysNGF2} following the procedure given in Kamrin and Henann \cite{kamrin2015} so that the model collapses the Froude number far away from flow threshold, in the limit $\chi \rightarrow 0$. {We also tried another version of that procedure replacing the quadratic term  with a cubic one as in the model of Lee and Yang \cite{lee2017}. Yet neither of these modified models performed} substantially better, with the no-slip solution still flowing significantly slower than the DEM data. This points to the boundary condition being the main culprit -- incidentally, the true {length scale $\delta$} at the bottom boundary is observed to increase with flow rate in the DEM data. With a velocity-dependent {$\delta$} that jumps from near zero in the arrested state to a large value in the flowing state, the NGF model could potentially produce correct predictions of both the transition stresses and the flowing regime. A boundary condition of this type could be formulated as a dynamical system governing the evolution of {$\delta$} in response to relevant driving quantities. As we note in the conclusion, however, formulating accurate and physically-justified fluidity boundary conditions remains a key open issue within NGF modeling, and such an endeavor is relegated to future work.


\section{Conclusions}

In this paper, we have studied the combined role of strain-rate weakening behavior and nonlocal effects in explaining key features of the hysteretic transition between solid-like and liquid-like behavior in dense granular materials as the applied stress is ramped up and down. These features include the hysteresis of the critical stresses at flow onset and arrest, the finite jump in velocity during flow onset, and the strengthening of the critical stresses with reducing system size. In a first part, we modified the nonlocal granular fluidity (NGF) model so that it reduces to a non-monotonic local rheology in homogeneous flows. Through numerical simulations of flow onset and arrest in planar shear with gravity using the modified NGF model, we demonstrated qualitatively that the inclusion of both nonlocal effects and non-monotonicity of the local rheology is essential to account for all three features mentioned above.

In a second part, we compared quantitatively predictions of the modified NGF model with DEM simulations of flow onset and arrest in various geometries. First, we calibrated the local parameters of the model using DEM simulations of homogeneous plane shear flow. In so doing, we highlighted the importance of calibrating the local critical stress for flow onset using stress-driven simulations, since measurements based on velocity-driven simulations are unreliable in the strain-rate weakening regime. The stress-driven simulations, however, exhibited large variability in the transition stresses between arrested and flowing regimes. Thus, we developed a criterion to extract a unique deterministic value, corresponding to the lowest observable critical stress, from a large number of repeated runs. We then compared predictions of the calibrated NGF model with stress-driven DEM simulations in planar shear with gravity and inclined plane configurations. In the former case, the model gave accurate predictions of both the transition between flowing and arrested states as well as the characteristics in the flowing regime. In the latter case, the accuracy of the model predictions was strongly affected by the choice of boundary conditions, with no single choice able to reproduce both the transition stresses as well as the average velocity in the flowing regime.

These results suggest that the NGF model generally leads to more accurate predictions when nonlocal effects are generated by the spatial dependence of the $\mu$ field, as is the case for planar shear with gravity, rather than by boundaries alone, as is the case for inclined plane flow. A possible explanation stems from Liu and Henann's \cite{liu2018} observation that an inhomogeneous $\mu$ field leads to much stronger size-dependent strengthening than boundary effects, making the accuracy of model predictions less reliant on the particular choice of fluidity boundary conditions when both mechanisms are present. Conversely, model predictions are much more sensitive to the specific type of boundary conditions when the latter are the sole source of nonlocal effects, which calls for better understanding of the interaction between flowing particles and the boundary. In fact, numerous experimental and computational studies have also underscored the sensitivity on wall roughness of transition stresses and velocity profiles in inclined plane flow \cite{rognon2002,silbert2002,goujon2003,weinhart2013}, plane shear flow without gravity \cite{schuhmacher2017}{, as well as annular shear flow \cite{fazelpour2021}}. Although progress has been made for flat frictional walls \cite{artoni2012,artoni2015}, the correct modeling of boundary conditions in the general case is still an open question whose resolution would benefit all nonlocal rheological models \cite{aranson2002,chaudhuri2012,liu2018}.

\section*{Conflicts of interest}
The authors have no conflicts to declare.

\appendix

\section{Coarse-graining methodology}
\label{app:CoarseGrainingProcedure}

In this appendix, we describe our coarse-graining procedure for extracting continuum velocity and stress fields from the particle-wise DEM data. The approach we follow was introduced in Zhang and Kamrin \cite{zhang2017}, building upon earlier work \cite{koval2009,andreotti2013,weinhart2013}. Since the geometries that we consider are homogeneous along the $x$-direction, the spatial averaging generates fields defined at discrete heights $z_k$, spaced $0.5d$ apart. For a given position $z_k$, we define the instantaneous velocity and stress fields as
\begin{align}
\mathbf{v}(z_k,t) &= \sum_{m = -M}^{M} w_m \bar{\mathbf{v}}(z_m,t), \label{eq:ContinuumVelocity} \\
\boldsymbol{\sigma}(z_k,t) &= \sum_{m = -M}^{M} w_m \bar{\boldsymbol{\sigma}}(z_m,t),
\end{align}
where $\bar{\mathbf{v}}$ and $\bar{\boldsymbol{\sigma}}$ are sublayer-wise velocity and stress averages at the heights $z_m = z_k + (W/2)(m/M)$, each weighted by the coefficient $w_m = \cos((\pi/2)(m/n))$. Following Zhang and Kamrin \cite{zhang2017}, we choose $W = 2d$ and $M = 5$. We now let $L_{im}$ denote the cross-sectional length between particle $i$ and the horizontal line at height $z_m$. The sublayer-wise instantaneous velocity is given by
\begin{equation}
\bar{\mathbf{v}}(z_m,t) = \frac{\sum_i L_{im} \mathbf{v}_i(t)}{\sum_i L_{im}},
\end{equation}
where $\mathbf{v}_i$ is the velocity of grain $i$. Then, the sublayer-wise instantaneous stress field is defined by
\begin{equation}
\bar{\boldsymbol{\sigma}}(z_m,t) = \frac{\sum_i L_{im} \boldsymbol{\sigma}_i(t)}{L},
\end{equation}
where $L$ is the domain length along the $x$-direction, and $\boldsymbol{\sigma}_i$ is the stress tensor associated with grain $i$. The latter consists of contact and kinetic contributions, and is given by
\begin{equation}
\boldsymbol{\sigma}_i(t) = \frac{1}{A_i} \sum_{i \neq j} \mathbf{f}_{ij}(t) \otimes \mathbf{r}_{ij}(t) + \frac{m_i}{A_i} \delta \mathbf{v}_i(t) \otimes \delta \mathbf{v}_i(t),
\end{equation}
where $A_i = \pi d_i^2/4$ and $m_i = \rho_s A_i$ are respectively the area and mass of grain $i$, $\mathbf{f}_{ij}$ is the contact force exerted on grain $i$ by grain $j$, and $\mathbf{r}_{ij}$ is the vector pointing from the center of grain $i$ to that of grain $j$. In the kinetic contribution, the velocity fluctuations are calculated as $\delta \mathbf{v}_i(t) = \mathbf{v}_i(t) - \mathbf{v}(z_i,t)$, where $\mathbf{v}(z_i,t)$ is the coarse-grained instantaneous velocity \eqref{eq:ContinuumVelocity} interpolated to the vertical position $z_i$ of grain $i$.

\section{Numerical discretization}
\label{app:NumericalDiscretization}

Here, we present our numerical discretization method for solving the fluidity equation \eqref{eq:HysNGF2} under a time-dependent applied stress. In the problems that we consider, the equation reduces to a one-dimensional PDE for $g(z,t)$ that is driven by a prescribed stress ratio function $\mu(z,t)$. The physics governing the time scale $t_0$ that appears in the fluidity equation are still unknown, and we simply assign a sufficiently small value $t_0 = 10^{-4} \,$s that it does not affect the dynamics of the solution. The spatial domain is discretized into $N = 100$ nodes, and the diffusion term is evaluated using second-order finite differences. The fluidity equation is then integrated in time using an implicit Euler scheme with a time step $\Delta t = 5 \cdot 10^{-4} \,$s, which we implement in MATLAB. Further, we artificially limit $g(z,t)$ to a minimum value of $10^{-2} \, \mathrm{s}^{-1}$ in order to avoid $g(z,t)$ reaching infinitesimally small values during the arrested portion of the stress ramp. Indeed, this would prevent $g(z,t)$ from growing sufficiently fast when flow onset should occur, as the applied stress is subsequently increased. We have verified that the floor value for $g(z,t)$ is small enough that it does not alter the observed transition stresses.

\section{Size effects in simple plane shear}
\label{app:SizeEffectsSimplePlaneShear}

\begin{figure}[htb]
 \centering
 \includegraphics[width=\columnwidth]{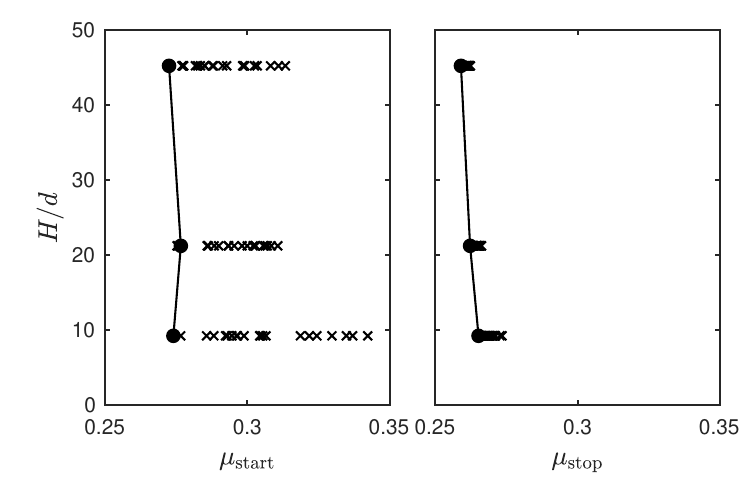}
 \caption{Critical stresses $\mu_\mathrm{start}$ and $\mu_\mathrm{stop}$ in simple plane shear versus true system height $H/d$, obtained from 20 different DEM runs. The crosses show the individual transition stresses pertaining to every run, while the filled circles represent the deterministic value extracted from the linear fit of the CDF.}
 \label{fig:PlaneShearSizeEffects}
\end{figure}

We report in Figure \ref{fig:PlaneShearSizeEffects} additional DEM results on the system-size dependence of the critical transition stresses $\mu_\mathrm{start}$ and $\mu_\mathrm{stop}$ in plane shear without gravity.
Mirroring our definition of $\mu_\mathrm{start}$ in Section \ref{sec:CalibrationSimplePlaneShear}, $\mu_\mathrm{stop}$ is defined as the observed $\mu_\mathrm{w}$ when $I_\mathrm{w}$ last falls below $10^{-3}$ during stress decrease. Shown are the individual transition stresses from 20 DEM runs (crosses) and the corresponding deterministic values (filled circles) extracted from the linear fit of the CDF according to the procedure outlined in Figure \ref{fig:Calibration}(c). We observe that the critical stresses are almost independent of system size, corroborating results from a previous DEM study\cite{chaudhuri2012}.

\section{Critical stresses from NGF}
\label{app:CriticalStressesNGF}

In this appendix, we explain how to obtain the critical starting and stopping stresses predicted by the NGF model without computing time-dependent solutions to a slowly varying applied stress, which are computationally intensive due to the required low rate of change of the applied stress to ensure quasi-steady conditions. Let the scalar $\bar{\mu}$ denote the amplitude of the applied stress ratio $\mu(z)$ throughout the domain -- for instance, $\bar{\mu}$ is $\mu_\mathrm{w}$ in the case of plane shear with gravity, or $\tan \theta$ in inclined plane. For a given geometry, implying a certain distribution of the stress ratio $\mu(z)$, we then rewrite the fluidity equation \eqref{eq:HysNGF2} as
\begin{equation}
t_0 \dot{g} = \mathcal{F}(g;\bar{\mu}),
\label{eq:NGFNonlinear}
\end{equation}
where the dependence on the magnitude of $\mu(z)$ has been explicitly denoted through $\bar{\mu}$. In the following, we will call $g_0$ any steady-state solution of \eqref{eq:NGFNonlinear}. 

At low $\bar{\mu}$, the arrested state $g_0 = 0$ is the only stable solution. Gradually increasing $\bar{\mu}$, flow onset occurs the moment the $g_0 = 0$ solution becomes unstable to small perturbations $g'$, which defines the critical starting stress $\bar{\mu}_\mathrm{start}$ \cite{aranson2002}. These small perturbations are governed by the linear equation
\begin{equation}
t_0 \dot{g}' = \mathcal{L}(g_0;\bar{\mu}) g',
\label{eq:NGFLinear}
\end{equation}
where $\mathcal{L}(g_0;\bar{\mu})$ -- the linearization (also called Fr\'echet derivative) of $\mathcal{F}(g;\bar{\mu})$ around $g_0$ -- acts on the perturbation $g'$ as
\begin{align}
\mathcal{L}(g_0;\bar{\mu}) g' &= A^2 d^2 \nabla^2 g' - \frac{(\mu_2-\mu_\mathrm{s})(\mu_\mathrm{s}-\mu)}{\mu_2-\mu} g' - 2 b \sqrt{\frac{m}{P}} \mu g_0 g' \nonumber \\
&\quad - \chi(g_0; \mu, P) g' - \frac{\partial \chi}{\partial g}(g_0; \mu, P) g_0 g'.
\end{align}
To evaluate whether perturbations grow or decay, we substitute $g' = \tilde{g}'(z) e^{\lambda t}$ into \eqref{eq:NGFLinear}, which leads to the eigenvalue problem
\begin{equation}
t_0 \lambda \tilde{g}'(z) = \mathcal{L}(g_0;\bar{\mu}) \tilde{g}'(z)
\label{eq:EigenvalueProblem}
\end{equation}
for the growth rate $\lambda$. This eigenvalue problem can be solved numerically by discretizing $\mathcal{L}(g_0;\bar{\mu})$ using a finite difference approximation, giving a spectrum of eigenvalues with the one having the largest real part, $\lambda_m$, dictating the overall rate of growth or decay of the perturbation. Setting $g_0 = 0$, one can perform repeatedly this calculation for increasing values of $\bar{\mu}$ until $Re\{\lambda_m\}$ becomes positive, at which point the arrested solution loses stability and $\bar{\mu} = \bar{\mu}_\mathrm{start}$.

Gradually decreasing $\bar{\mu}$ from a value above $\bar{\mu}_\mathrm{start}$, flow arrest occurs the moment \eqref{eq:NGFNonlinear} ceases to admit a nonzero steady-state solution $g_0$, which defines the critical stopping stress $\bar{\mu}_\mathrm{stop}$. To check whether that is true at a given value of $\bar{\mu}$, it suffices to perform Newton-Raphson iterations to find $g_0$. Starting from an initial guess $g^0$, the algorithm performs at each step $n$ the update $g^{n+1} = g^n + \omega \Delta^n$, where $0 < \omega < 1$ is a relaxation parameter and the step direction $\Delta^n$ is given through the linear system
\begin{equation}
\mathcal{L}(g^n;\bar{\mu}) \Delta^n = - \mathcal{F}(g^n;\bar{\mu}).
\label{eq:LinearSystem}
\end{equation}
We stop the iterations when the norm of $\mathcal{F}(g^n;\bar{\mu})$ falls under a specified threshold, indicating that $g^n$ has converged to a steady-state solution $g_0$ of \eqref{eq:NGFNonlinear}. Thus, our strategy to find $\bar{\mu}_\mathrm{stop}$ goes as follows. We start with a value of $\bar{\mu}$ above $\bar{\mu}_\mathrm{start}$, for which we are guaranteed a nonzero $g_0$ solution. We compute the latter by letting $g$ reach steady-state in the time-dependent solver. Then, we repeatedly compute $g_0$ for incrementally decreasing values of $\bar{\mu}$ through Newton-Raphson iterations, using at each step level of $\bar{\mu}$ the converged solution $g_0$ from the previous step as an initial guess. At some point the Newton-Raphson iterations will suddenly converge to the arrested $g_0 = 0$ solution, indicating that $\bar{\mu}$ has reached $\bar{\mu}_\mathrm{stop}$.

Both the eigenvalue problem \eqref{eq:EigenvalueProblem} and linear system \eqref{eq:LinearSystem} are implemented in MATLAB borrowing the same grid and discretized differential operators used in the time-dependent solver. In the Newton-Raphson iterations, we use $\omega = 0.5$ to balance stability and speed of convergence. Previous studies \cite{kamrin2015,liu2018} have shown that for some geometries and boundary conditions, there exist analytical or semi-analytical solutions for the growth rate $\lambda$ and thus the threshold $\mu_\mathrm{start}$. However, such solutions are much harder to obtain for $\mu_\mathrm{stop}$, despite partial progress in that direction on a $I$-gradient model applied to the inclined plane scenario \cite{lee2017}. Therefore, we limit ourselves in this paper to the numerical methodology that we have outlined above, noting that it is computationally very efficient -- the starting and stopping curves in Figures \ref{fig:PlaneShearGravity} and \ref{fig:InclinedPlane} were calculated in a few minutes on a laptop.

\section{Hysteresis size dependence}
\label{app:HysteresisSizeDependence}

\begin{figure}[htb]
 \centering
 \includegraphics[width=\columnwidth]{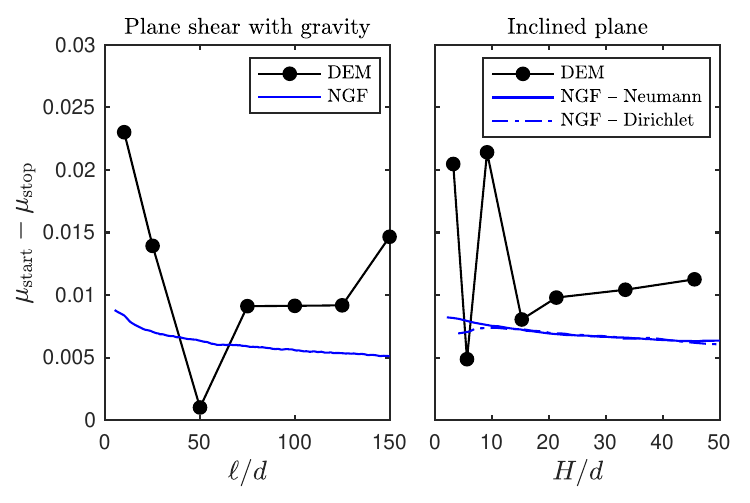}
 \caption{Critical stress difference $\mu_\mathrm{start} - \mu_\mathrm{stop}$ versus dimensionless loading length scale $\ell/d$ for plane shear with gravity, and versus dimensionless layer height $H/d$ for inclined plane. The DEM data (filled circles) and NGF data (blue lines) is the same as in Figures \ref{fig:PlaneShearGravity}(c,f) and \ref{fig:InclinedPlane}(c,f).}
 \label{fig:CriticalStressDifference}
\end{figure}

Here, we investigate the system-size dependence of the hysteresis amplitude, measured by the difference of the starting and stopping critical stress ratios, based on the DEM simulations and NGF model predictions reported in Figures \ref{fig:PlaneShearGravity} and \ref{fig:InclinedPlane} in Sections \ref{sec:PlaneShearGravity} and \ref{sec:InclinedPlane}. Figure \ref{fig:CriticalStressDifference} displays $\mu_\mathrm{start} - \mu_\mathrm{stop}$ as a function of the dimensionless loading length scale $\ell/d$ for plane shear with gravity data, and as a function of the dimensionless layer height $H/d$ for inclined plane data.
The DEM and NGF data are within a comparable range and suggest a weak effect of system size on the hysteresis amplitude.



\balance


\bibliography{bibliography} 
\bibliographystyle{rsc} 

\end{document}